\documentclass{article}

\usepackage{arxiv}
\usepackage[utf8]{inputenc} 
\usepackage[T1]{fontenc}    
\usepackage{hyperref}       
\hypersetup{
    colorlinks=true,
    linkcolor=black,
    citecolor=black,
    urlcolor=black
    }
\usepackage{url}            
\usepackage{booktabs}       
\usepackage{amsfonts}       
\usepackage{amsmath}
\usepackage{amssymb}
\usepackage{nicefrac}       
\usepackage{microtype}      
\usepackage[capitalize]{cleveref}       
\usepackage{lipsum}         
\usepackage{graphicx}
\usepackage[numbers,sort&compress]{natbib}
\usepackage{doi}
\usepackage{authblk}
\usepackage[linesnumbered,ruled,vlined]{algorithm2e}
\usepackage{bm}

\usepackage{caption}
\usepackage{subcaption}
\usepackage{multirow}
\usepackage{lineno}

\usepackage[most]{tcolorbox}
\usepackage{xcolor}

\newtcolorbox{reviewercomment}{
  colback=blue!8,      
  colframe=blue!8,     
  boxrule=0pt,         
  arc=0pt,             
  left=10pt,
  right=10pt,
  top=6pt,
  bottom=6pt,
  breakable             
}
\newcommand{\mbf}[1]{\mathbf{#1}}

\DeclareMathOperator{\red}{red}
\DeclareMathOperator{\unq}{unq}
\DeclareMathOperator{\syn}{syn}
\DeclareMathOperator{\res}{res}

\DeclareMathOperator{\sgn}{sgn}

\begin{document}

\title{Redundancy Maximization as a Principle of Associative Memory Learning in Hopfield Networks}
\date{October 2025}

\renewcommand\Authfont{\bfseries}
\setlength{\affilsep}{0em}
\author[1,$\dagger$]{Mark Blümel}
\author[2,1,$\dagger$]{Andreas C. Schneider}
\author[2,1,$\dagger$]{Valentin Neuhaus}
\author[2,3,1]{David A. Ehrlich}
\author[4]{Marcel Graetz}
\author[3,1]{Michael~Wibral}
\author[3,1]{Abdullah Makkeh}
\author[1,2]{Viola Priesemann}

\affil[1]{Complex Systems Theory, Max Planck Institute for Dynamics and Self-Organization, Göttingen, Germany}
\affil[2]{Faculty of Physics, Institute for the Dynamics of Complex Systems, University of Göttingen}
\affil[3]{Göttingen Campus Institute for Dynamics of Biological Networks, University of Göttingen, Göttingen, Germany}
\affil[4]{Champalimaud Centre for the Unknown, Lisbon, Portugal}
\affil[$\dagger$]{These authors share first authorship.}
\affil[ ]{\texttt{viola.priesemann@ds.mpg.de}}
\maketitle

\begin{abstract}
Associative memory, traditionally modeled by Hopfield networks, enables the retrieval of previously stored patterns from partial or noisy cues. Yet, the local computational principles which are required to enable this function remain incompletely understood. To formally characterize the local information processing in such systems, we employ a recent extension of information theory---Partial Information Decomposition (PID). PID decomposes the contribution of different inputs to an output into \emph{unique} information from each input, \emph{redundant} information across inputs, and \emph{synergistic} information that emerges from combining different inputs. Applying this framework to individual neurons in classical Hopfield networks we find that below the memory capacity, the information in a neuron's activity is characterized by high redundancy between the external pattern input and the internal recurrent input, while synergy and unique information are close to zero until the memory capacity is surpassed and performance drops steeply. Inspired by this observation, we use redundancy maximization at each neuron as an information-theoretic learning goal. This dramatically increases the network's memory capacity to $1.59$, a more than tenfold improvement over the $0.14$ capacity of classical Hopfield networks, and also outperforming recent state-of-the-art implementations of Hopfield networks. Overall, this work establishes redundancy maximization as a new design principle for associative memories and opens pathways for new associative memory models based on information-theoretic goals.

\end{abstract}

\section{Introduction}

Associative memory---the ability to retrieve patterns from noisy or partial inputs---is a fundamental brain function, enabling the retrieval of memories from imperfect sensory stimuli. This type of content-addressable memory can be modeled by recurrent neural networks called ``Hopfield networks'', for which their inventor John Hopfield was recognized with the Nobel Prize in physics in 2024~\citep{hopfield1982neural, nobel2024physics}. Recently, continuous-valued extensions of Hopfield networks have also found renewed application in machine learning~\citep{ramsauer2020hopfield}.

Despite decades of development since their first introduction in 1982, the principles underlying associative memory formation remain incompletely understood. Originally, Hopfield networks were trained using the biologically-inspired Hebbian learning rule based on firing coincidences. Since then, new learning rules have been introduced that display improved memory capacity and stability ~\citep{hillar2012efficient, Tolmachev_2020}. Nevertheless, a key question remains: Is there an underlying principle that governs the formation of associative memory? And, if so, can it be exploited directly to improve performance?

To answer these questions, we propose to analyze Hopfield networks from an information processing perspective. Hopfield networks store patterns as attractors of their neural dynamics, created by training the network's weights using the patterns as a teaching signal. How the information of the recurrent dynamics and the teaching signal together predicts the neuron's firing thus becomes pivotal to the network's performance. Describing this relation in an abstract, implementation-independent manner can be achieved using the framework of information theory~\cite{shannon, coverthomas}. 
In particular, the mutual information quantifies general statistical dependencies between two variables, enabling the quantification of the amount of information in a neuron's output that is contained in the input from other neurons (\emph{recurrent input}) or in the teaching signal (\emph{target input}). 
However, using only mutual information it is impossible to differentiate \emph{how} that information is contributed in unique, redundant, or synergistic ways. We can differentiate these input contributions using a recent extension of information theory called Partial Information Decomposition (PID, \cite{williams2010nonnegative}). 

In this work, we use PID as a tool to describe the individual neuron's function in a Hopfield network in both an analytic and constructive manner. The analysis of classical Hebbian Hopfield networks reveals that neurons show high redundancy in their firing between recurrent and teaching inputs. Building on this insight, we then use PID to construct novel local goal functions which optimize for redundancy and other information processing goals directly.

Information theory and the concept of PID have been used to analyze information processing in artificial neural networks in the past~\citep{shwartz2017opening,geiger2021information,ehrlich2023a, gutknecht2025}. Furthermore, it has been shown that ``infomorphic'' neurons which train by directly optimizing a PID goal function are in principle capable of solving tasks from different learning paradigms~\cite{makkeh2023general} and have been demonstrated to achieve high performance and local interpretability for supervised learning tasks in particular~\cite{schneider2024should}.

The main contributions of this work are (i) an information-theoretic analysis of a classical Hebbian Hopfield network which reveals that redundancy between recurrent input and target (teaching signal) dominates the neurons' output when operating below capacity, (ii) the construction of ``infomorphic'' associative memory neurons that directly maximize a local PID goal function, building on and improving the proof of concept provided by \citet{makkeh2023general}, and (iii) experimental results showcasing that infomorphic neurons that maximize redundancy significantly outperform the original Hebbian learning rule and other previous learning methods.

\section{Methodological Approach}
\label{sec:Background}

\begin{figure}
    \centering
    \includegraphics[width=1.0\linewidth]{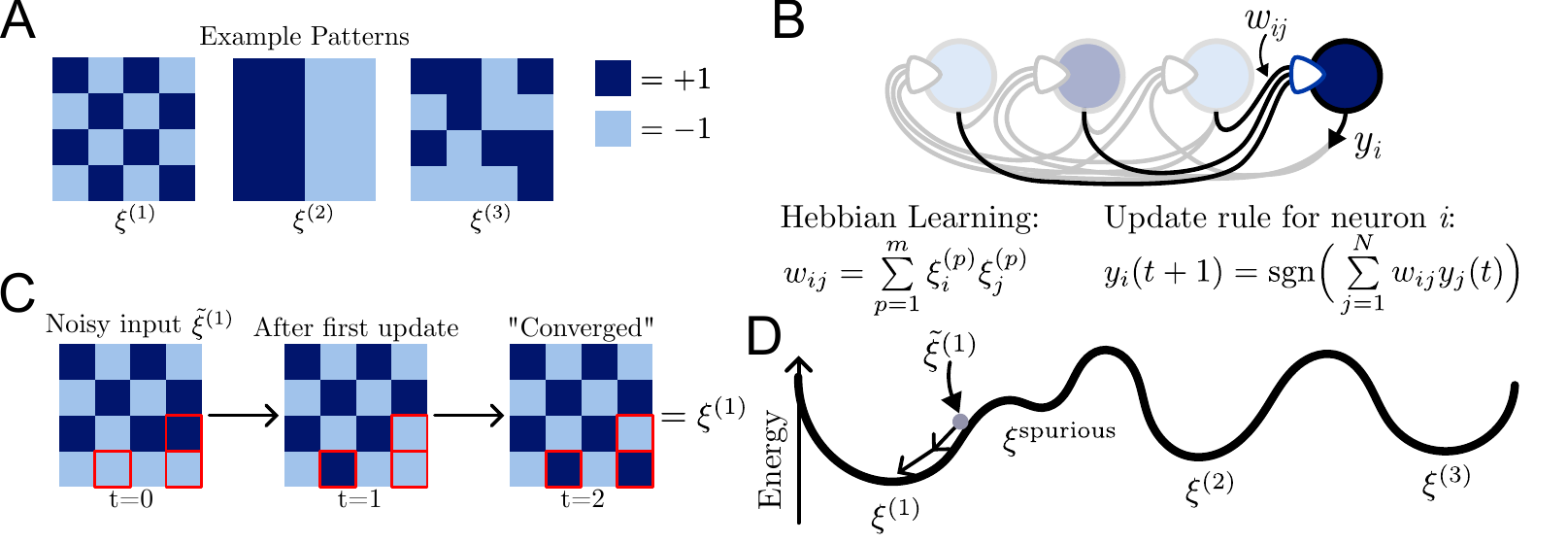}
    \caption{ \textbf{Hopfield networks store patterns as attractors and retrieve them by iteratively updating noisy inputs.} \textbf{A:} Three example patterns that could be stored in a Hopfield network (vectors $\xi^{(p)}$ re-formatted to 2-D for visual clarity). Based on the patterns, the weights of the network (\textbf{B}) can be directly computed. \textbf{C:} The neurons are initialized in a state that is a corrupted version of a pattern (here $\xi^{(1)}$ with corruptions marked in red) and the network updates its states based on the given update rule. If successful, the network converges to the original pattern. \textbf{D}: The stored patterns $\xi^{(p)}$ become minima in the energy landscape of the network. If the model starts close to a stored pattern, it will converge back to its minimum energy state and recovers the pattern. However, if the capacity of the network is exceeded, i.e.\ the minima become too close in the energy landscape, new local minima can form, resulting in so-called ``spurious memories''. This energy landscape view holds only for symmetric Hopfield networks (i.e., where $w_{ij} = w_{ji}$).}
    \label{fig:HopfieldCartoon}
\end{figure}

\subsection{Hopfield Networks}
\label{ssec:HopfieldNetworks}
A Hopfield network consists of $N$ recurrently connected neurons. Initialized with a noisy or incomplete version of a stored pattern, the network iteratively updates neuron states to minimize an energy function, converging to a stable state that corresponds to the stored pattern. This allows for pattern completion and error correction through memory retrieval (see \autoref{fig:HopfieldCartoon}).

The neurons are updated based on the activity of other neurons in the network. Given a network state vector $\{y_i(t)\}_i$ at time $t$, the state of neuron $i$ at the next timestep is computed as
\begin{equation}
\label{eq:NeuronUpdate}
y_i(t+1) = \sgn\left( r_i \right),
\end{equation}
where $\sgn(x)$ denotes the sign function and the recurrent input $r_i$ is defined as a weighted sum of the activities of all neurons $y_j$ as
\begin{equation*}
    r_i = \sum_{j=1}^N w_{ij} y_j(t).
\end{equation*}

The originally proposed training method for the weights $\{w_{ij}\}_{ij}$ was Hebbian learning. In this learning rule, neurons which co-activate as part of the same pattern become more strongly connected, making the patterns attractors in the energy landscape. For bipolar patterns $\boldsymbol \xi^{(p)}$ (see \autoref{fig:HopfieldCartoon}\textbf{A} for example patterns) with pattern index $p$ (i.e., with pattern elements $ \xi_i^{(p)} = \pm1$) the resulting weights using the Hebbian learning rule are given by \citep{hopfield1984neurons}
\begin{equation}
    \label{eq:HebbBinary}
    w_{ij} = \sum_{p=1}^m \xi_i^{(p)}\xi_j^{(p)},
\end{equation}
where $m$ is the total number of patterns. There are typically no self-connections in the network and therefore $w_{ii} = 0$. 
When training on uncorrelated patterns, the amount of memories (patterns) $m_c$ that a Hopfield network can successfully store scales linearly with network size $N$. The corresponding memory capacity is defined as $\alpha_c = \frac{m_c}{N}$ and depends on the learning method. The Hebbian rule has a capacity of $\alpha_\mathrm{H} \approx 0.14$ \citep{amit1985storing}, lower than the theoretical upper bound of $\alpha_\mathrm{c} = 2$~\citep{cover1965geometrical, gardner1988space}.

\subsection{Partial Information Decomposition}
\label{ssec:pid}
To understand how individual neurons contribute to associative memory function, we investigate the information processing at a single neuron. From an information-theoretic viewpoint, each neuron can be viewed as a channel that receives information about the outputs of other neurons via the recurrent input $R$ and produces a bipolar output signal $Y$. The total entropy of a neuron's output $H(Y)$ can be decomposed into two parts, namely the information $I(Y:R)$ explained by $R$, and the residual entropy $H(Y\mid R)$, such that $H(Y) = I(Y:R)+H(Y\mid R)$.

\begin{figure}[t]
    \centering
    \includegraphics{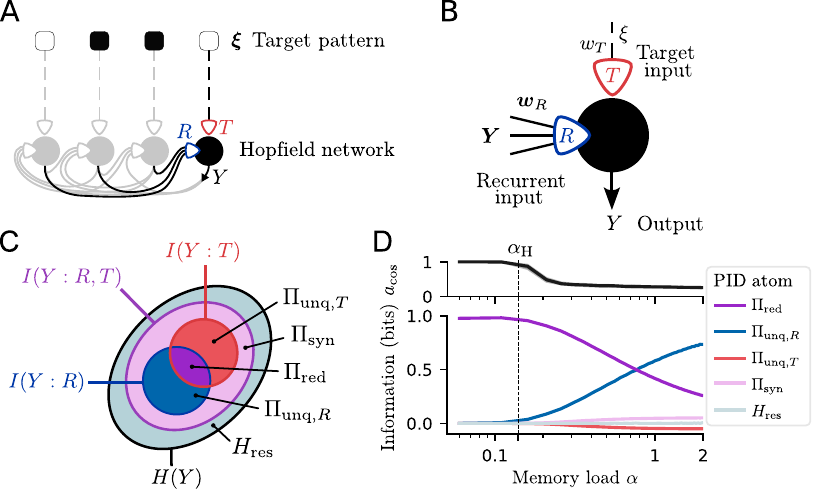}
    \caption{\textbf{For classical Hopfield networks trained with Hebbian learning, redundant information between target and recurrent input coincides with successful memory storage.} \textbf{A:}  Schematic of the analysis set-up for Hopfield networks. To measure how information is represented, each neuron is compared to a non-driving target input $T$ that provides the ground-truth pattern, in addition to its recurrent input $R$.
    \textbf{B:} Each neuron in the Hopfield network aggregates its recurrent inputs and produces an output $Y$. The neurons are initialized in the target state $\xi$.
    \textbf{C:} Partial information decomposition (PID) separates the entropy of the output $Y$ into five parts: Unique information (provided by only one of the two inputs), redundant (shared by both inputs), synergistic (emerging only from the combination of inputs) and residual entropy (not explained by the inputs). 
    \textbf{D:} The PID profile as a function of memory load $\alpha$. Below the networks memory capacity ($\alpha_\mathrm{H} \approx 0.14$, indicated by dashed black line), the redundancy $\Pi_{\red}$ is high. Above capacity, as recall fails, redundancy collapses and is replaced by unique information from the recurrent input. The accuracy of recall is shown in black. The PID profiles show the median of $20$ network initializations, with values first averaged across all neurons. The accuracy curve is the median of the $20$ initializations. Shaded areas indicate the central $90\,\%$ percentile of values. Results are from a network with $N=500$ neurons to minimize finite-size effects (see \autoref{app:finite}). A more technical and analytical discussion of the results, including a reproduction of panel \textbf{D} using other PID measures, can be found in Appendix \ref{app:HebbianRedundancy}.
    }
    \label{fig:hebb}
\end{figure}

To evaluate whether this output information is relevant to the memory task, we introduce an auxiliary input variable $T$ that corresponds to the targeted output for a given pattern and acts as a teaching signal for the neuron~(\autoref{fig:hebb}\textbf{A}). While the total information which the recurrent and the new target input together hold about the neuron output $Y$ is quantified by the joint mutual information $I(Y:R,T)$, classical information theory cannot explain \emph{how} this information is provided by the two source variables $R$ and $T$: Some parts of the neuron's firing might be explainable \emph{uniquely} by the recurrent input (denoted by $\Pi_{\unq, {R}}$) and thus be unrelated to the target input, or uniquely by the target input, i.e., relevant but not encoded in $R$ ($\Pi_{\unq, T}$). Other parts of the output information may be carried \emph{redundantly} ($\Pi_{\red}$) in both sources, meaning they are both relevant and encoded in $R$, while yet others may be carried \emph{synergistically} ($\Pi_{\syn}$), meaning that both information sources are necessary to uncover this piece of information. Enumerating and quantifying these information \emph{atoms} is the subject of Partial Information Decomposition (PID)~\citep{williams2010nonnegative, gutknecht2021bits}.

The four PID atoms are related to the three classical mutual information terms via the so-called consistency equations~\cite{williams2010nonnegative} (see ~\autoref{fig:hebb}\textbf{C}.)
\begin{equation}
    \begin{aligned}
    \label{eq:consistencyeq}
    I(Y:R) &= \Pi_{\unq,R} + \Pi_{\red} , \\
    I(Y:T) &= \Pi_{\unq,T} + \Pi_{\red} , \\
    I(Y:R,T) &= \Pi_{\unq,R} + \Pi_{\unq,T} + \Pi_{\red} + \Pi_{\syn}.
    \end{aligned}
\end{equation}

This set of linear equations is underdetermined and leaves open a degree of freedom in the values of atoms $\Pi$. To resolve this underdetermination, an additional quantity needs to be defined, which is usually a measure for redundancy. Throughout the literature, a plethora of different redundancy measures have been devised, which fulfill distinct requirements and have different operational interpretations \cite [e.g.][and references therein]{lizier2018information}. Throughout this work, we use the $I_\cap^\mathrm{sx}$ measure introduced by \citet{makkeh2021introducing}, due to its differentiability, which is essential for optimization (see \autoref{ssec:training}). Using PID now allows for a more fine-grained decomposition of the entropy of $Y$ into five components
\begin{equation}
    H(Y) =  \Pi_{\unq,R} + \Pi_{\unq,T} +\Pi_{\red}  + \Pi_{\syn} + H_\text{res},
\end{equation}
where $H_\text{res} = H(Y\mid R,T)$ denotes the residual entropy of the output $Y$ not explained by either source variable $R$ or $T$ (see Appendix \ref{app:calculating_atoms} for explicit calculations of the information atoms).

\subsection{Information-theoretic Learning}
\label{ssec:InfomorphicNets}
PID provides a valuable framework to analyze the information processing of an individual neuron in a Hopfield network. However, to investigate whether this information processing view provides a sufficient description of the neurons' functions, PID can also be employed to build novel ``infomorphic neurons'' \citep{makkeh2023general}, which optimize a PID-based goal function directly.

During evaluation, these infomorphic networks operate analogously to classical Hebbian Hopfield networks with the neuron's activity given by \autoref{eq:NeuronUpdate}. The network state is initialized in a target pattern, and neurons synchronously update their output based on both inputs until convergence or up to 100 time steps using the same update rule (\autoref{eq:NeuronUpdate}).

For training only, the true pattern element is introduced as an additional target input $T$ to each neuron to guide the learning. The recurrent dynamics are updated only once, this time utilizing a sigmoid activation function to produce a firing probability and stochastically assigning $+1$ or $-1$ to $Y$ according to this probability. After this, the joint input distribution is estimated using a differentiable ``soft'' binning approach (see \autoref{app:Binning}), enabling computation of PID atoms using the discrete $I^\text{sx}_\cap$ PID measure. Given these PID atoms, a general infomorphic objective function of the form
\begin{equation}
\label{eq:loss}
    G = \gamma_{\unq,R} \Pi_{\unq,R} + \gamma_{\unq,T} \Pi_{\unq,T} + \gamma_{\red} \Pi_{\red} + \gamma_{\syn} \Pi_{\syn} + \gamma_{\res} H_{\res},
\end{equation}
can be evaluated, where $\gamma_i$ are fixed parameters that control the maximization and minimization of individual information atoms. Finally, the weights of each infomorphic neuron are locally updated by gradient ascent on $G$ using automatic differentiation.

Performance is assessed using cosine similarity between recalled and target patterns, with accuracy quantified both as an average over all patterns ($a_\mathrm{cos}$) and via a threshold measure ($a_\theta$) to obtain the fraction of patterns recalled with above $\theta = 95 \%$ accuracy. The memory capacity ($\alpha_c$) is determined by evaluating accuracy at different memory loads using different random seeds, and stability is quantified by measuring the maximum fraction of corrupted bits $f_\mathrm{max}$ in the initial network state. More details can be found in \autoref{app:extended_methods}.

\section{Results}
\subsection{Redundancy in Hebbian Learning}
\label{sec:analysis}

To understand the information-theoretic footprint of Hebbian Hopfield networks, we first investigate a network with $500$ neurons trained using Hebbian learning. When computing the PID between the recurrent inputs $r_i$ to the neuron and the target patterns $t_i$ introduced as an additional variable with no direct influence on the output (see \autoref{fig:hebb}\textbf{A},\textbf{B}), one can see that the information contributions of the inputs change for different memory loads $\alpha$: For memory loads below and close to the memory capacity of $\alpha_\mathrm{H} \approx 0.14$, redundant information is the largest information contribution (see \autoref{fig:hebb}\textbf{D}), indicating that high redundant information $\Pi_{\red}$ coincides with successful learning. 
Above the memory capacity, redundant information begins to fall while the unique information $\Pi_{\unq,R}$ begins to rise, indicating that the neurons start to encode more non-relevant information from the recurrent inputs. In Appendix \ref{app:HebbianRedundancy} we provide a formal connection between high redundancy and high recall accuracy in Hebbian Hopfield networks as well as empirical results using other PID measures.

Overall, the results suggest that high redundant information may play an important role in successful associative memory function.

\subsection{Redundancy as a Computational Goal}
\label{sec:redRule}
To evaluate whether the information-theoretic description found in \autoref{sec:analysis} is sufficient to describe associative memory function, this section introduces a constructive approach to show how maximization of redundancy can be used as a computational goal for associative memory learning and shows that it outperforms the classical Hopfield network.

\begin{figure}[t]
    \centering
    \includegraphics{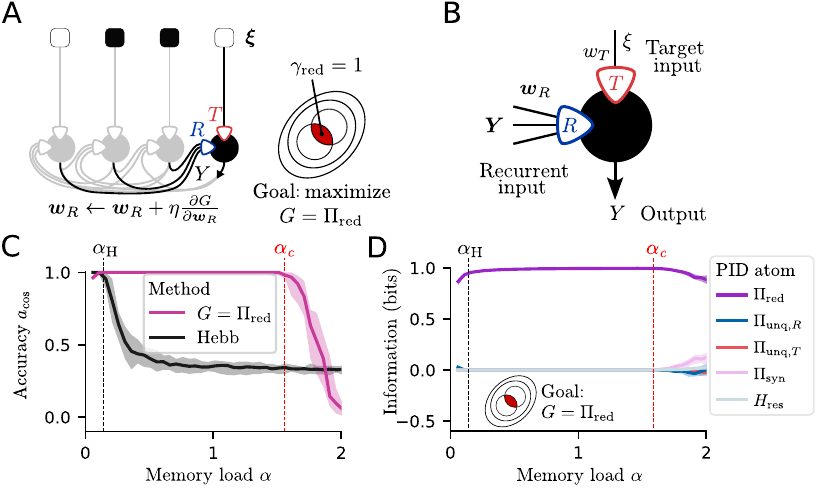}
    \caption{\textbf{Redundancy maximization between recurrent connections and a target is a sufficient principle for memorization in Hopfield networks, achieving a memory capacity of $\alpha_\mathrm{c}^{\red} \approx 1.59$.} \textbf{A:} Schematic of the infomorphic Hopfield model. During training of the infomorphic Hopfield network, the recurrent connections $\boldsymbol{w}_R$ are updated using gradient ascent on the goal function $G=\Pi_{\red}$.
    \textbf{B:} Each neuron in the infomorphic Hopfield network aggregates its input into two compartments---the recurrent input $R$ and the target input $T$. Based on this, they stochastically produce an output $Y$.
    \textbf{C:} Recall accuracy as a function of memory load $\alpha$ for a network with $100$ neurons. Redundancy maximization achieves a memory capacity of $\alpha_\mathrm{c}^{\red} = 1.59 \ [1.56, 1.61]$, far exceeding the Hebbian capacity of $\alpha_\mathrm{H} \approx 0.14$ (capacities marked by dashed lines).
    \textbf{D:} The PID profile shows the mean information atoms per neuron as a function of memory load $\alpha$. Redundancy dominates below capacity $\alpha <\alpha_\mathrm{c}^{\red}$, then falls off as the capacity is crossed and the other atoms become non-zero. 
    In \textbf{C}, the curve shows the median calculated across $20$ network initializations. In \textbf{D}, values are first averaged across all neurons, and the curve then shows the median of these averages across the initializations. For both panels, the shaded area represents the central $90 \%$ of the data, spanning the 5th to the 95th percentile.}
    \label{fig:capacity}
\end{figure}

To this end, we employ ``infomorphic neurons'' in a training procedure described in \autoref{ssec:training} (see also \autoref{fig:capacity}\textbf{A} and \textbf{B}). Motivated by the results of the previous section, we first use the simple neuron goal function $G = \Pi_{\red}$, i.e., we train each neuron to maximize the redundant information between its inputs about the output. 

To evaluate the performance of these infomorphic Hopfield networks, we train networks with $N=100$ neurons for different numbers of patterns and apply the testing procedure described in \autoref{ssec:testing}. We find that the network memorizes all patterns up to a memory load of $\alpha_\mathrm{c}^{\red} = 1.59 \ [1.56, 1.61]$ followed by a sharp drop in recall accuracy above this memory capacity (see \autoref{fig:capacity}\textbf{C}). The angular brackets indicate bootstrapped $95 \%$ confidence intervals of the capacity $\alpha_c$.

The transition above $\alpha_\mathrm{c}^{\red}$ can be explained by performing the same PID analysis as in \autoref{sec:analysis}. As shown in \autoref{fig:capacity}\textbf{D}, redundancy is uniformly high below $\alpha_\mathrm{c}^{\red}$ apart from some finite size effects for low numbers of patterns discussed in \autoref{app:finite}. However, when the threshold is crossed, the information encoded by the neurons is no longer purely redundant. Instead, as the neurons fail to encode only task-relevant redundant information, small amounts of other information atoms, particularly synergistic information $\Pi_{\syn}$, start to arise.

\subsection{Mutual Information Goals}
\label{sec:classicInfo}
\begin{figure}[t]
    \centering
    \includegraphics{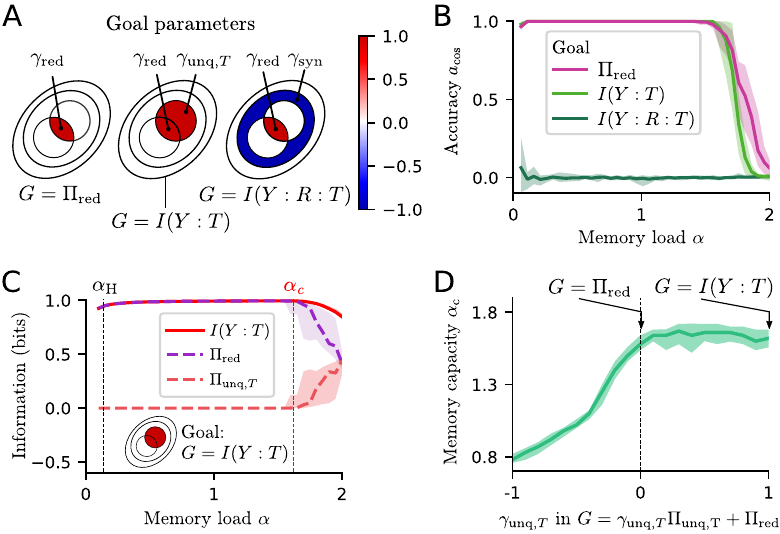}
    \caption{\textbf{Infomorphic Hopfield networks trained with a classical mutual information goal achieve high memory capacity by implicitly maximizing redundancy.}
    \textbf{A:} The redundancy goal and two alternative learning objectives based on classical information theory: Maximizing the mutual information between output and target, $G=I(Y:T)$, and maximizing the co-information, $G= I(Y : R :T)$. 
    \textbf{B:} A performance comparison of the two alternative goals as a function of memory load shows that maximizing mutual information ($G=I(Y:T)$) achieves a very similar capacity as maximizing redundancy ($G=\Pi_{\red}$) alone. In contrast, maximizing co-information ($G= I(Y : R :T)$) fails to store any patterns. 
    \textbf{C:} Information profiles for the successful $G=I(Y:T)$ goal. Beyond its memory capacity $\alpha_{c}$, mutual information $I(Y:T)$ stays high, but redundancy ($\Pi_{\red}$) falls and is replaced by unique information from the target ($\Pi_{\unq,T}$).
    \textbf{D:} The memory capacity $\alpha_\mathrm{c}$ is shown as a function of $\gamma_{\mathrm{unq},T}$ in the goal $G = \gamma_{\mathrm{unq},T} \Pi_{\mathrm{unq},T} + \Pi_{\red}$. While positive values of $\gamma_{\unq,T}$ have no strong effect on the memory capacity, negative values are detrimental. 
    The curves in \textbf{B} and \textbf{D} show the median across 20 network initializations while in \textbf{C}, values are first averaged across all neurons. In all panels, the shaded area represents the central $90 \%$ of the data.
    }
    \label{fig:classical}
\end{figure}

Besides optimizing redundancy or other PID-based goal functions directly, the infomorphic approach also allows us to compare the former to goal functions based on classical information theory alone. To this end, we consider two main candidate goal functions (\autoref{fig:classical}\textbf{A}), with two further variants discussed below: (i) the mutual information between the target inputs and the neuron's output $I(Y:T)$ and (ii) the co-information, $I(Y: R : T)$, which is sometimes used as a classical measure of the tradeoff between redundancy and synergy~\cite{kay2011coherent}. Co-information is formally defined as
\begin{equation}
    \begin{aligned}
        I(Y:R:T) &= I(Y:R)+I(Y:T)-I(Y:R,T)\\
        &= \Pi_{\red}-\Pi_{\syn}.
    \end{aligned}
\end{equation}

As before, we first evaluate the capacity of the networks trained with the two classical goal functions (\autoref{fig:classical}\textbf{B}). While maximizing co-information fails to store any patterns, maximizing target mutual information $I(Y:T)$ achieves very similar performance to maximizing redundancy and reaches a memory capacity of $\alpha_{\mathrm{c}} = 1.62 \ [1.61, 1.64]$. In addition, the joint mutual information $I(Y:T,R)$ reaches a capacity of $\alpha_{\mathrm{c}} = 0.85 \ [0.84,0.86]$, while the recurrent mutual information $I(Y:R)$ fails to learn any patterns.

The high performance of the mutual information optimization can be understood by looking at the PID profile shown in \autoref{fig:classical}: The mutual information $I(Y:T)$ can be decomposed into $\Pi_{\red}$ and $\Pi_{\unq,T}$, i.e., into a redundant and a unique contribution. Below the maximum capacity $\alpha_\mathrm{c}$, the mutual information is maximized only by maximizing the redundant information $\Pi_{\red}$ between the two inputs, explaining the similar performance to optimizing this redundancy alone. Above that point, the mutual information remains high, but redundancy decreases, matched by an increase in unique information of the target input $\Pi_{\unq,T}$. The corresponding decrease in accuracy suggests that the mutual information goal works only as long as it results in a maximization of redundancy, further suggesting redundancy as the underlying principle of Hopfield memory function.
The effect of varying the unique contribution $\gamma_{\unq,T}$ in the goal function  $G = \gamma_{\unq,T} \Pi_{\unq,T} + \Pi_{\red}$ indicates that additional maximization of target unique information hardly impacts the memory capacity while negative values lead to a decrease in performance (see \autoref{fig:classical}\textbf{D}).

\subsection{Improving the Capacity of the Infomorphic Goal}
\label{sec:optRedRule}
\begin{figure}[t]
    \centering
    \includegraphics{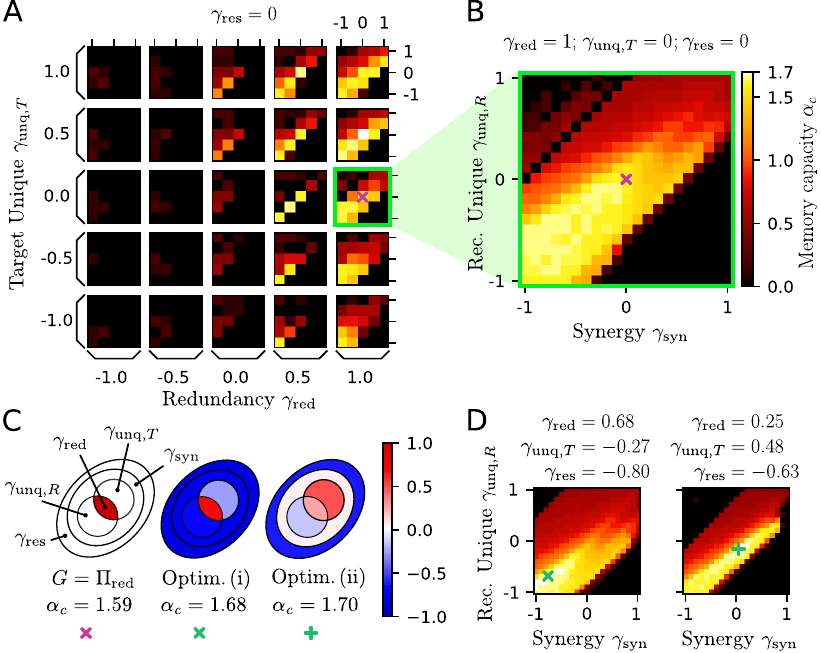}
    \caption{\textbf{Hyperparameter optimization reveals composite information goals which outperform redundancy maximization.}
    \textbf{A:} The memory capacity landscape $\alpha_{c}$ as a function of the goal parameters $\gamma_i$ reveals the performance across large parts of the parameter space. The goal parameters for redundancy $\gamma_{\red}$ and target unique information $\gamma_{\unq,T}$ vary on the outer axes, while the goal parameters for synergy $\gamma_{\syn}$ and recurrent unique information $\gamma_{\unq,R}$ vary on the inner axes. The remaining goal parameter for residual entropy $\gamma_{\res}$ is fixed at 0.
    \textbf{B:} A more detailed subspace of \textbf{A} reveals the area around the heuristic goal. Here, $\gamma_{\unq,T} = 0$ and $\gamma_{\red} = 1$ while $\gamma_{\unq,R}$ and $\gamma_{\syn}$ are varied. Suppressing both $\Pi_{\unq,R}$ and $\Pi_{\syn}$ slightly improves the capacity above redundancy maximization.
    \textbf{C:} A direct optimization of the capacity on the full goal space reveals new goals. The heuristic redundancy goal as well as two of the best performing goals are illustrated. The optimized goals reach a memory capacity of  $\alpha^\mathrm{(i)}_\mathrm{c} = 1.68 \ [1.65, 1.72]$  and $\alpha^\mathrm{(ii)}_\mathrm{c} = 1.7 \ [1.66,1.71]$. The exact goal parameters and the results of the optimizations are listed in \autoref{app:Optimizationresults}.
    \textbf{D:} As in \textbf{B}, the panel shows two slices of the landscape around the optimized goals. The landscapes show a similar structure with distinct local optima.}
    \label{fig:regimes}
\end{figure}
While optimizing redundancy has proven sufficient for achieving high associative memory performance, it remains open whether this simple intuitive goal is optimal or whether admixtures of other PID atoms can improve performance even further. To investigate this, we performed a parameter sweep over the $\gamma$ parameters in the general goal function $G=\gamma_{\unq,R}\Pi_{\unq,R}+\gamma_{\unq,T}\Pi_{\unq,T}+\gamma_{\red}\Pi_{\red}+\gamma_{\syn}\Pi_{\syn}$, with a fixed parameter for residual entropy $\gamma_{\res}=0$. We then measured the network capacities for the different optimization goals (see \autoref{fig:regimes}).

The resulting performance landscape reveals that the highest capacities are indeed achieved in proximity of the heuristic redundancy goal $G = \Pi_{\red}$ (\autoref{fig:regimes}\textbf{A}), consistent with the intuition that in order to memorize a pattern, each neuron's recurrent activation must align with its target activation. The performance remains mostly stable for positive additions of $\Pi_{\unq,T}$, as long as the redundant contribution dominates. Suppressing the target-unrelated information of each neuron by choosing negative parameters for both recurrent unique information and synergy, $\gamma_{\unq,R}$ and $\gamma_{\syn}$, appears to slightly increase performance, but only if both parameters are approximately equal (\autoref{fig:regimes}\textbf{B}). In line with this result, the capacity decreases when either or both of them are positive.

Similar to the approach of~\citet{schneider2024should}, we further used a hyperparameter optimization technique to find optimal hyperparameter combinations. We used the CMA-ES sampler \citep{hansen2001completely} to find the goal parameters that maximize the capacity of the model.
Two of the best goals found with this approach reach capacities of $\alpha^\mathrm{(i)}_\mathrm{c} = 1.68 \  [1.65, 1.72]$ and $\alpha^\mathrm{(ii)}_\mathrm{c} = 1.7 \ [1.66,1.71]$, slightly outperforming the heuristic redundancy maximization ($\alpha_\mathrm{c}^{\red} = 1.59 \ [1.56, 1.61]$) (see \autoref{fig:regimes}\textbf{C}). The two goal functions are
\begin{align}
    G^{(i)} &= -0.27\,\Pi_{\unq,T} -0.68\,\Pi_{\unq,R} + 0.68\,\Pi_{\red} - 0.77\,\Pi_{\syn}  - 0.80\,H_{\res} \text{\quad and}\\
    G^{(ii)} &= 0.48\,\Pi_{\unq,T} - 0.16\,\Pi_{\unq,R} + 0.25\,\Pi_{\red} + 0.04\,\Pi_{\syn} - 0.63\,H_{\res}.
\end{align}
While both goals maximize redundancy, the first suppresses all other contributions and the second only suppresses $\Pi_{\unq,R}$ and $H_{\res}$ while it maximizes $\Pi_{\unq,T}$, similar to the mutual information goal $I(Y:T)$. The results of the other optimizations can be found in \autoref{app:Optimizationresults}. The landscapes around the optimized goals have a similar qualitative structure in the subspace of $\gamma_{\unq, R}$ and $\gamma_{\red}$, while the optima are at different values (see \autoref{fig:regimes}\textbf{D}).

\subsection{Performance Comparison}
\label{sec:comparison}
\begin{figure}[t]
    \centering
    \includegraphics{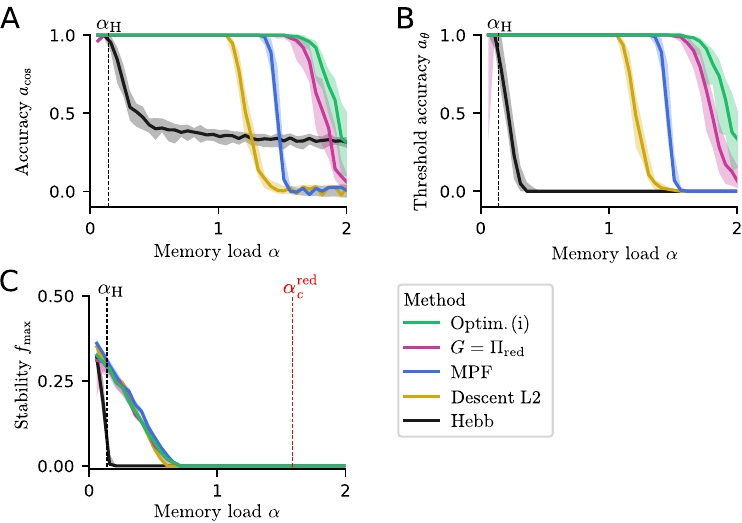}
    \caption{
    \textbf{In terms of memory capacity, the infomorphic approach outperforms other high-performance learning methods.}
    \textbf{A:} Comparison of the performance of the Hebbian learning rule, our heuristic and optimized infomorphic goals, and two high-performance methods, the descent L2 method \cite{Tolmachev_2020, DiedrichOpper} and the minimum probability flow (MPF)~\cite{hillar2012efficient}. Using cosine similarity ($a_\mathrm{cos}$) as the accuracy metric, both infomorphic goals outperform all other methods. Above capacity, Hebbian learning displays spurious memories~\cite{amit1985spinglass}.
    \textbf{B:} Same as in \textbf{A}, but accuracy is measured using a strict threshold ($a_\theta$ with $\theta=0.95$). This metric reveals that the Hebbian learning rule is unable to fully reconstruct patterns above its capacity $\alpha > \alpha_\mathrm{H}$ as its accuracy falls to zero.
    \textbf{C:} The stability of memory retrieval is shown as a function of memory load $\alpha$. The figure is adapted from \citet{Tolmachev_2020}, with us adding MPF and the infomorphic goals. The infomorphic method is on par with the other goals, although it was not even designed to optimize stability. The curves in \textbf{A}-\textbf{D} show the median values across 20 network initializations, and the shaded areas denote the corresponding central $90 \%$ percentile.}
    \label{fig:comparison}
\end{figure}
Infomorphic Hopfield networks achieve competitive performance compared to other state-of-the-art learning rules for Hopfield networks. To demonstrate this, we selected two learning methods which are also based on goal functions---the minimum probability flow goal (MPF)~\cite{hillar2012efficient} and descending the L2 norm (Descent L2)~\cite{Tolmachev_2020, DiedrichOpper}). Initially introduced by \citet{sohl2011new}, MPF demonstrates that by minimizing the probability of entering neighboring states of the
pattern states $\xi$, many patterns can be encoded in a robust way. Our approach differs from MPF by introducing local goal functions at the level of each neuron and by explicitly connecting the probabilistic approach to information theory.
\citet{Tolmachev_2020} demonstrate that learning in a Hopfield network can often be implemented as descent of different loss functions that describe the stability of the training patterns. They compare different norms, revealing that minimizing the L2 norm between the pattern and the next evolved time step achieves high memory capacity.
Both MPF and Descent L2 reach state-of-the-art capacities above the Hebbian learning rule, but are outperformed by the infomorphic redundancy maximization (see \autoref{fig:comparison}\textbf{A,B}).

In addition, we also evaluate the stability of the memories encoded in different networks (\autoref{fig:comparison}\textbf{C}). The stability is measured by introducing progressively more noise $f$ into the patterns and testing whether recall of memories is still successful. The area under the curve $f_\mathrm{max}$ is a proxy for the overall stability of memories and indicates the point up to which $95\,\%$ of the pattern elements are reconstructed. For more details, see \autoref{ssec:training}.
The redundancy goal's profile for stability follows an overall trajectory similar to that of the other learning methods. Note that above $\alpha \approx 0.8$ stability is reduced to almost zero for all methods. 

\section{Discussion}

\paragraph{Summary}
This work presents a new perspective on associative memories by examining the information-processing at the scale of individual neurons. Using PID, we quantify how recurrent and target inputs predict neuronal output. We demonstrate that effective pattern storage requires the recurrent and target inputs to redundantly determine firing, highlighting redundancy as a key local information-processing principle underlying associative memory function. Analyzing classical Hopfield networks (\autoref{sec:analysis}), we find that below the memory capacity, neurons predominantly process information redundantly, while above capacity, redundancy degrades and unique recurrent contributions increase, indicating a loss of task-relevant encoding (\autoref{fig:hebb}\textbf{D}).

This connection is not coincidental: as we show analytically in Appendix~\ref{app:HebbianRedundancy}, when using a Heaviside activation function, successful recall requires the conditional input distributions $p(R\mid T{=}\pm w_T)$ to have non-overlapping support across the decision boundary---a condition that is mathematically equivalent to maximal redundancy regardless of how the weights are obtained. We corroborate this finding with a closed-form derivation for Hebbian networks that reproduces the empirical profile of \autoref{fig:hebb}\textbf{D}, and holds across alternative PID measures and network sizes (Appendices~\ref{app:BinningStability} and~\ref{app:finite}).

Building on this insight, we construct infomorphic Hopfield networks where each neuron directly maximizes redundant information between its inputs. These networks significantly outperform classical Hebbian Hopfield networks in memory capacity and stability (\autoref{sec:redRule}). Further optimizing the redundancy-based objective enhances performance by suppressing irrelevant information contributions (\autoref{sec:optRedRule}, \autoref{fig:regimes}\textbf{C}). Using only classical information-theoretic measures, we find that maximizing mutual information between target pattern elements and neuron output yields comparable performance to PID-based models, which can, however, be ascribed to an implicit optimization of redundancy (\autoref{fig:classical}).

Across varying memory loads, the observed correspondence between high redundancy and recall accuracy supports our claim that it underlies successful associative memory learning (\autoref{fig:classical}\textbf{C}). Finally, benchmarking against state-of-the-art methods confirms that infomorphic Hopfield networks achieve superior memory loads, up to $\alpha_\mathrm{c} = 1.7 \ [1.66, 1.71]$, while maintaining competitive stability (\autoref{fig:comparison}).

\paragraph{Related Work}

The approach of constructing local information-theoretic goal functions to perform associative memory learning has previously been suggested by~\citet{makkeh2023general}. In this work, we improve upon these results by replacing the complex custom activation function used by \citet{makkeh2023general} by a sign function during testing, which is replaced by a differentiable sigmoid for training, as well as a novel ``soft'' binning procedure, and demonstrate that $G=\Pi_{\red}$ and the classical goal $I(Y:T)$ both achieve a memory capacity that is fourfold the capacity reported in~\cite{makkeh2023general}. Furthermore, we show that a direct optimization of the goal parameters further increases the memory capacity.

In addition, information theory has previously been employed to analyze Hopfield networks. Notably, the mutual information of the target pattern and the internal state $I(\boldsymbol \xi: \boldsymbol Y)$ has been prominently utilized. \citet{dominguez1998self} initially established $I(\boldsymbol \xi: \boldsymbol Y)$ as a suitable measure for evaluating the quality of pattern retrieval. \citet{dominguez2004mutual, dominguez2009structured} subsequently demonstrated that $I(\boldsymbol \xi: \boldsymbol Y)$ was optimal for a sparsely connected network, which ultimately enhanced the memory capacity. Additionally, information-theoretic concepts have been used to train (deep) neural networks \citep{linsker1988application, kay2011coherent, makkeh2023general}. Previous research has demonstrated the significance of maximizing $I(\boldsymbol \xi: \boldsymbol Y)$ in enhancing the memory capacity, which in this work we explain by its implicit maximization of redundancy.

\paragraph{Limitations and Outlook}
Our results demonstrate that maximizing redundancy directly can significantly improve memory capacity for Hopfield networks. However, promoting redundancy alone does not suffice to reach the theoretical optimum $\alpha_\mathrm{c} = 2$ of twice the number of neurons established by \citet{cover1965geometrical}. Beyond classical associative memories, it has been shown that this bound can be exponentially exceeded by dense associative memories, which incorporate a layered architecture to improve performance~\cite{krotov2016dense, krotov2023new}. Similarly, \citet{schneider2024should} constructed infomorphic networks with multiple layers, achieving substantial improvements in supervised learning tasks compared to single-layer networks. These findings open an intriguing avenue for future research: leveraging the framework of infomorphic networks to identify the information-processing mechanisms required to realize dense associative memories with hidden units. We hypothesize that the general principle of redundancy will continue to play a central role.

Furthermore, since the infomorphic update rule requires the estimation of the full joint probabilities, the approach is computationally expensive and biologically implausible on its own. However, the main advantage of the infomorphic approach lies in its flexible and interpretable goal functions. In future research, approximations to these learning rules may be devised which more implicitly achieve the principle of redundancy maximization.

A further important direction concerns correlated patterns. While it has been shown in classical Hebbian Hopfield networks that spatial correlation is detrimental to storage capacity~\cite{correlationcapacity}, we expect that infomorphic Hopfield networks could perform better in this setting, since infomorphic networks can better differentiate which target pattern elements are truly informative of the output. Promising first results in this direction are shown in \autoref{app:Correlations}.

Another interesting research direction is to apply the principles of memory found here to complex objectives. For example, curiosity has been theorized as the seeking of novel patterns or information based on memorized patterns~\cite{schmidhuber}. Infomorphic networks could reproduce this behavior by following a mixed goal, maximizing redundancy for memorization and unique information from the target for exploration.

In summary, Partial Information Decomposition (PID) and infomorphic networks form a powerful pair of tools for analysis and construction of associative memory networks. Together, they offer a new level of interpretability at the local level that can be used to assess the strengths and weaknesses of existing learning methods and to inspire the design of novel network architectures with state-of-the-art performance in associative memory tasks.
\section*{Code Availability}
The code to reproduce the results of this work is submitted with the paper and will be made public at the date of publication.
\section*{Acknowledgements}
We would like to thank the members of the Priesemann and Wibral groups for valuable feedback and discussions. M.B. was supported by the Max Planck Society. A.S., V.N. and V.P. were funded via the MBExC by the Deutsche Forschungsgemeinschaft (DFG, German Research Foundation) under Germany’s Excellence Strategy-EXC 2067/1-390729940. A.S., V.N., M.W., and V.P. were supported and funded by the DFG – GRK2906 – project number 502807174 and acknowledge support from the Max Planck Society. V.N. was partly supported by the Else Kröner Fresenius Foundation via the Else Kröner Fresenius Center for Optogenetic Therapies. D.A.E. and M.W. were supported by a funding from the Ministry for Science and Education of Lower Saxony and the Volkswagen Foundation through the “Niedersächsisches Vorab” under the program “Big Data in den Lebenswissenschaften” – project “Deep learning techniques for association studies of transcriptome and systems dynamics in tissue morphogenesis”. M.W. and A.M. are employed at the Campus Institute for Dynamics of Biological Networks (CIDBN) funded by the Volkswagenstiftung. M.W., V.P. and A.M. received funding from the DFG via the SFB 1528 “Cognition of Interaction” - project-ID 454648639. M.W. was supported by the flagship science initiative of the European Commission’s Future and Emerging Technologies program under the Human Brain project, HBP-SP3.1-SGA1-T3.6.1. M.G. received a PhD scholarship by the Champalimaud Foundation.

\bibliographystyle{unsrtnat}
\bibliography{references}  

\newpage

\appendix
\section{Extended Methods}
\label{app:extended_methods}

\subsection{Calculating PID atoms}
\label{app:calculating_atoms}
In the case of two source variables, the calculation of all PID atoms follows from the three consistency equations (see \autoref{eq:consistencyeq})
\begin{equation*}
    \begin{aligned}
    I(Y:R) &= \Pi_{\unq,R} + \Pi_{\red} , \\
    I(Y:T) &= \Pi_{\unq,T} + \Pi_{\red} , \\
    I(Y:R,T) &= \Pi_{\unq,R} + \Pi_{\unq,T} + \Pi_{\red} + \Pi_{\syn}.
    \end{aligned}
\end{equation*}
and one additional axiom. In our case, we define redundant information using the $I_\cap^\mathrm{sx}$ measure introduced by \citet{makkeh2021introducing}, obtaining
\begin{equation}
    \begin{aligned}
    \label{eq:Isx}
    \Pi_{\red} := I_\cap^\mathrm{sx}(Y:R;T)  &:= \sum_{r,t,y} p(r,t,y) i_\cap^\mathrm{sx}(y:r;t)\\ &:= \sum_{r,t,y} p(r,t,y) \log \frac{p(y,r\cup t)}{p(y)p(r\cup t)}\\ &:= \sum_{r,t,y} p(r,t,y) \log \frac{p(y,r) + p(y,t) - p(y,r,t)}{p(y)[p(r)+p(t)-p(r,t)]}.
    \end{aligned}
\end{equation}
Given this definition, the other atoms are calculated via Moebius inversion:
\begin{equation}
    \begin{aligned}
    \label{eq:Moebius}
    \Pi_{\unq,R} &= I(Y:R) - \Pi_{\red} , \\
    \Pi_{\unq,T} &= I(Y:T) - \Pi_{\red} , \\
    \Pi_{\syn} &= I(Y:R,T) - \Pi_{\unq,R} - \Pi_{\unq,T} - \Pi_{\red}.
    \end{aligned}
\end{equation}

\subsection{Infomorphic Hopfield Network Set-up}
The architecture of infomorphic Hopfield networks is similar to that of the classical Hopfield network introduced in \autoref{ssec:HopfieldNetworks}. 
The number of infomorphic neurons $N$ corresponds to the number of pattern elements of the target patterns to be stored in the recurrent weights of the network. During training, each neuron receives two different inputs. The first input is a weighted sum of the outputs of the recurrently connected neurons
\begin{equation}
\label{eq:lateralInput}
    r_i = \sum_j^N w_{ij}^R y_j.
\end{equation}
The second, target input $T$ is given by the (weighted) element of the target pattern that the neuron encodes, i.e.,
\begin{equation}
    t_i = w_{i}^T \xi_i^{(p)}.
\end{equation}
Given these two inputs, the probability of neuron $i$ returning state $+1$ is given by
\begin{align}
    p(y_i=+1) = \sigma\left(r_i + t_i\right),
    \label{eq:stochastic}
\end{align}
with $\sigma$ indicating the sigmoid function, while the probability of state $-1$ is given by $1-p(y_i=+1)$.

\subsection{Training}
\label{ssec:training}
By treating the inputs and the output as random variables, we apply PID to optimize specific information-processing objectives. Before training, the network and optimizer are initialized according to \autoref{tab:params} in \autoref{app:params}.

The training procedure itself begins by initializing the network state in the target pattern $\xi^{(p)}$ and presenting single elements $\xi^{(p)}_i$ of the same pattern via the target input $t$ as teaching signal to each corresponding neuron $i$. The network is then run for one time step in a synchronous manner to obtain the recurrent inputs $r$ and the output probabilities $p(Y = y \mid R = r, T = t)$ according to \autoref{eq:stochastic}. Over all patterns, we estimate the empirical joint probability mass functions $p(R = r, T = t)$ of $R$ and $T$ for each neuron using the (soft) plug-in binning method described in \autoref{app:Binning}.

Given these probability estimates, the PID terms are computed as described in Appendix \ref{app:calculating_atoms}. Then, the objective function $G$ is evaluated according to \autoref{eq:loss}. Finally, the recurrent weights $\boldsymbol{w}_R$ of each neuron are updated via gradient ascent on $G$, using automatic differentiation. While in principle the target weight $w_T$ of each neuron could also be updated, in practice a constant weight of order 1 resulted in fastest learning and best performance. Specifically, we chose a constant $w_T = 2.3$.

\subsection{Testing}
\label{ssec:testing}
To evaluate the performance of the trained network, we use a procedure similar to Hopfield. First, the neurons are initialized in the pattern $\xi^{(p)}$. As before, the recurrent inputs are given by \autoref{eq:lateralInput}, but the neurons do not receive target input signals. The bipolar output of the neurons is then given by
\begin{equation}
    y_i(t+1) = \sgn(r(t)) 
\end{equation}
which is identical to the update rule in \autoref{eq:NeuronUpdate}. This step is repeated until the network converges to a fixed point or limit cycle, or for a maximum of 100 iterations. The network’s performance is then evaluated based on the cosine similarity between the final network state $\mbf y(t_f)$ and the target pattern $\boldsymbol \xi_i^{(p)}$. We use two measures to aggregate the individual similarities and arrive at an overall accuracy score.
The first is the mean accuracy
\begin{align}
    a_{\mathrm cos} = \frac{1}{P}\sum_p \langle(\mbf y(t_f),\boldsymbol\xi^{(p)}) \rangle,
\end{align}
where we average alignment both across neurons and across patterns. The second is the threshold accuracy $a_\theta$
\begin{align}
    a_\theta = \frac{1}{P}\sum_p \mathcal{H}(\langle(\mbf y(t_f),\boldsymbol \xi^{(p)}) \rangle -\theta),
\end{align}
where $\mathcal{H}$ is the Heaviside function and $\theta$ the threshold parameter, which we set to $\theta =0.95$. For this measure to be non-zero, at least some patterns must be restored with high accuracy. 

To arrive at the memory capacity of a given learning method, we validate it on 20 different seeds. On each seed, we train networks with memory loads $\alpha$ increasing with a step size of $0.02$. We take as the memory capacity the value $\alpha_\mathrm{c}$, at which for all $\alpha < \alpha_\mathrm{c}$, the network encodes memories with $a_\mathrm{cos}(\alpha)>0.95$ (excluding finite size effects). Unless mentioned otherwise, we take the median over all validation seeds to arrive at a memory capacity. 

\subsubsection{Stability}
In addition to estimating the performance, we also estimate the stability of the stored memories. We test the stability by introducing an increasing fraction of flips $f$ to a pattern $\xi^{(p)}$ to initialize the network state and then measuring the accuracy $a_\mathrm{cos}(f)$. The stability $f_\mathrm{max}$ is given by the maximum fraction of flips under which the original memory can still be recovered with an accuracy $a_\mathrm{cos}(f_{max}) \geq \epsilon $ of at least $\epsilon=0.95$. By varying the memory load $\alpha$ before measuring the stability, we can obtain a full stability profile $f_\mathrm{max}(\alpha)$. The area under the curve $f_\mathrm{max}(\alpha)$ can be viewed as a proxy for the overall level of stability achieved by a given rule. The choice of parameters is taken as in \citet{Tolmachev_2020}.

\subsection{Model Parameters}
\label{app:params}
In \autoref{tab:params}, we explain the parameters of infomorphic networks and their values that were used during training unless specified otherwise. 
\begin{table}[ht]
    \caption{The model parameters.}
    \label{tab:params}
    \centering
    \begin{tabular}{ccc}
        \toprule
        Parameter & Definition/Meaning & Value \\
        \midrule 
        $N$ & number of neurons & 100 \\
         $\boldsymbol{w}_{T}^{\text{Hebb}}$ & initialization of the target weights used in Hebbian setup  & $10^{-10} \cdot \boldsymbol{1}_N$ \\
        $ \boldsymbol{w}_{T}^{\text{Info}}$ & initialization of the target weights used in infomorphic setup & $2.3 \cdot \boldsymbol{1}_N$ \\
        $\boldsymbol{w}_{R}$ & initialization of the recurrent weights & $\sim\mathcal{N}\left(0, \lambda_r  \sqrt{2/N}\right)$\\
        $\lambda_r$ & initial scale of recurrent weights & $1 \cdot 10^{-3}$ \\
        optimizer & algorithm to maximize $G$ &Adam \\
        $\eta$ &learning rate  &0.05 \\
        epochs & number of complete passes of the entire input patterns &5000 \\
        reps & number of consecutive times the pattern is presented &1 \\
        $n_t$ & number of bins in target direction& 2 \\
        $n_r$ & number of bins in recurrent direction & 60 \\
        $\sigma_t$& target kernel width as fraction of bin width & $1 \cdot 10^{-6}$\\
        $\sigma_r$& recurrent kernel width as fraction of bin width & $0.5$\\
        padding & how far the binning range extends beyond the data & 1\\
        sequential & True if states are updated sequentially &False \\
        $N_{\text{iter}}$ & maximum iterations during testing &$100$ \\
        $\theta$ & testing threshold for $a_\theta$ & 0.95 \\
        \bottomrule
    \end{tabular}
\end{table}
\section{Pseudocode for Training}

\SetAlFnt{\fontsize{10.1pt}{12pt}\selectfont} 
\SetAlCapFnt{\large}
\SetAlCapNameFnt{\large}
\SetArgSty{textnormal}
\SetKwComment{Comment}{/* }{ */}
\SetAlgorithmName{Function}{function}{List of Functions}

\begin{algorithm}
\caption{TrainHebbianHopfieldModel\label{alg:Hebbian}}
\KwIn{data}
\KwOut{trained model}
INITIALIZE model\;
INITIALIZE model.neuron\_weights $\gets$ zero\_matrix\;
\ForEach{\text{pattern in data}}{
    model.neuron\_weights $\gets$ model.neuron\_weights + outer(pattern, pattern)\;
}
\Return{model}
\end{algorithm}

\begin{algorithm}
\caption{TrainInfomorphicHopfieldModel}\label{alg:infomorphic_training}
\KwIn{data, num\_epochs, goal\_params}
\KwOut{trained model}
INITIALIZE model\;
\ForEach{ epoch in range(num\_epochs)}{
    INITIALIZE model\_outputs;
    
    \ForEach{\text{pattern in data}}{
        INITIALIZE network\_state $\gets$ pattern\;
        network\_state $\gets$ model.forward\_network(t=pattern, r=network\_state)\;
        model\_outputs.append(network\_state)\;
    }
    \ForEach{neuron in model}{
        TrainInfomorphicNeuron(neuron, goal\_params, y=output\_state[neuron], t=data, r=model\_outputs)\;
    }
    
}
\Return{model}
\end{algorithm}

\begin{algorithm}
\caption{TrainInfomorphicNeuron\label{alg:traininfoneuron}}
\KwIn{neuron, goal\_params, $y$, $r$, $t$}
\KwOut{None}
BIN continuous values $r$ in 60 and $t$ in 2 equally sized bins\;
COUNT occurrences of tuples $(r, t)$\;
COMPUTE empirical probability masses $p(r, t)$\;
EVALUATE conditional probabilities $p(y\mid r, t)$ from the neurons\;
CONSTRUCT full joint probability mass function $p(y, r, t) = p(r, t)p(y \mid r, t)$\;
isx\_redundancies $\gets$ ComputeIsxRedundancies($p(y, r, t)$)\;
pid\_atoms $\gets$ ComputePIDAtoms(isx\_redundancies)\;
goal $\leftarrow$ scalar\_product(goal\_params, pid\_atoms)\;
PERFORM autograd to maximize goal\;
UPDATE neuron.weights\;
\end{algorithm}

\begin{algorithm}
    \caption{ComputeIsxRedundancies}
    \KwIn{Joint probability mass function $p(y, r, t)$}
    \KwOut{Isx Redundancy Measure Values}
    \ForEach{antichain $\beta\in \big\{\{\{1\},\{2\}\}, \{\{1\}\}, \{\{2\}\}, \{\{1,2\}\} \big\}$}{
        COMPUTE conditional probability mass functions $p(Y=y\mid\bigvee_{\bm b \in \beta}\bigwedge_{i \in \bm b} S_i=s_i)$\;
        COMPUTE marginal probability mass function $p(Y=y)$\;
        $I^\mathrm{sx}_\cap(Y:S_\beta) \gets \sum_{y, r, t} p(Y=y, R=r, T=t) \log_2 \frac{p(Y=y \mid\bigvee_{\bm b \in \beta}\bigwedge_{i \in \bm b} S_i=s_i)}{p(Y=y)}$\;
        }
    \Return{$I^\mathrm{sx}_\cap(Y:S_\beta)$ for all antichains $\beta$}
\end{algorithm}

\section{Optimization Results}
Each run of the optimization algorithm takes place on a single seed, producing a set of best goals which cluster close together. We have taken one candidate goal function from each of the 10 optimization results and verified them on a set of 20 different seeds. The results are depicted in \autoref{fig:regimes}. Out of these ten candidate goals, the best two were chosen for \autoref{fig:comparison}. 
\label{app:Optimizationresults}
\begin{table}[ht]
\caption{Estimates for the memory capacities shown in the main text, including the bootstrapped $95\%$ confidence intervals. Additionally, the goals and capacities for other local optima found during the optimization procedure described in \autoref{sec:optRedRule} are shown.}
    \centering
    \begin{tabular}{ccc||ccccc}
    \hline
\multirow{2}{*}{Goal Name} & \multicolumn{2}{c||}{Capacity}            & \multicolumn{5}{c}{Goal Parameter}                                                          \\
                      & Median & \multicolumn{1}{l||}{$95\%$ CI} & $\gamma_{\unq,T}$ & $\gamma_{\unq,R}$ & $\gamma_{\red}$ & $\gamma_{\syn}$ & $\gamma_{\res}$ \\
 
    \hline
    Redundancy $\Pi_\mathrm{red}$& 1.59 & [1.56, 1.61] & 0 & 0 & 1 & 0 & 0 \\
    Target information $I(Y:T)$ & 1.62& [1.61, 1.64] & 1 & 0 & 1 & 0 & 0 \\
    \hline
    &1.60 & [1.58, 1.62] & 0.52 & -0.41 & 0.94 & -0.31 & 0.57 \\
    &1.64 & [1.62, 1.65] & 0.43 & -0.51 & 0.72 & -0.65 & -0.05 \\
    &1.66 & [1.63, 1.67] & 0.26 & -0.38 & 0.78 & -0.4 & -0.34 \\
    &1.66 & [1.64, 1.68] & 0.18 & -0.36 & 0.68 & -0.51 & -0.17 \\ 
    &1.66 & [1.62, 1.69] & 0.41 & -0.08 & 0.8 & -0.17 & 0.04 \\ 
    &1.68 & [1.65, 1.7] & 0.06 & -0.16 & 0.78 & -0.08 & -0.37 \\ 
    &1.68 & [1.66, 1.7] & 0.84 & 0.16 & 0.92 & 0.36 & -0.46 \\ 
    &1.7 & [1.66, 1.72] & 0.26 & -0.38 & 0.78 & -0.4 & -0.34 \\ 
    \hline
    Optim. (i) & 1.68 & [1.65, 1.72] & -0.27 & -0.68 & 0.68 & -0.77 & -0.80 \\
    Optim. (ii) & 1.7& [1.66, 1.71] & 0.48 & -0.16 & 0.25 & 0.04 & -0.63 \\
    \end{tabular}
    
    \label{tab:optimization}
\end{table}

\section{Binning}
\label{app:Binning}
\subsection{Differentiable Binning}
\begin{figure}[t]
    \centering
    \includegraphics{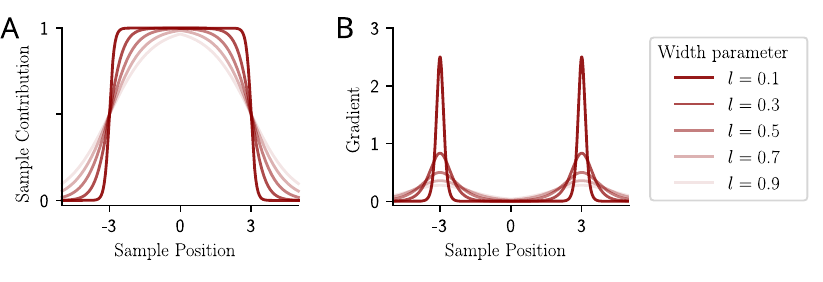}
    \caption{\textbf{In the soft binning procedure, the histograms are smoothed at the bin edges resulting in non-zero gradients of the histogram.} \textbf{A:} The contribution of a single sample to a bin covering the range $[-3,3]$. The contribution scales with the distance to the center of the bin. \textbf{B:} The corresponding magnitude of gradients to \textbf{A}. If $l$ approaches zero, we converge to hard bin edges and the gradients become delta peaks.}
    \label{fig:sigmoid_binning}
\end{figure}

The PID terms are estimated using a discrete PID measure due to the lack of a differentiable PID measure for mixed discrete-continuous variables~\cite{ehrlich2024partial}. Thus, the continuous-valued inputs $r$ and $t$ are binned to allow estimating a discrete probability distribution $p(r,t)$. Since this operation is not differentiable, previous approaches to estimating the PID terms assumed that $p(r,t)$ is constant with respect to the weights $w_R$ and $\boldsymbol{w}_T$, resorting to estimating $p(r,t)$ from a static histogram~\cite{kay1994information}. Taking into consideration the small changes in $p(r,t)$ with respect to the weights, however, might improve the PID estimates. Therefore, we implemented a kernel-based 'soft binning' approach that smooths the histogram at the bin edges providing estimates of $p(r,t)$ gradients (\autoref{fig:sigmoid_binning}). 

\subsection{Soft Binning Procedure}
The soft binning procedure is composed of three steps. First, for each bin the distances $d$ between its edges and the coordinates of every sample $(r,t)$ in all dimensions is calculated. Then, each bin is assigned weights according to a kernel $K$ for every sample. The choice of $K$ and its properties are discussed in~\autoref{apx:kernel}. Finally, the bin weights obtained for every sample are normalized, so that the total count of the histogram adds up to the number of samples.

\subsection{Kernel choice\label{apx:kernel}}
The kernel $K(d)$ should be differentiable, monotonically fall to zero as distance of the bin to the sample increases and approach a delta function as its width parameter decreases. We chose a sigmoid kernel $K(d)= \Pi_{i \in \{x,y\}}\sigma((c_i/2-d)/l)$ that satisfies all of these constraints, where $c_x$ and $c_y$ are the the histogram sizes and $l$ is the width parameter. As $l$ tends to zero, the non-differentiable binning values are recovered (\autoref{fig:sigmoid_binning}).

Note that the normalization step in the binning procedure ensures that each sample has an equal impact on the total results. Otherwise, especially when $l$ is smaller than half a bin size, samples that by chance lie closer to the center of a bin are more impactful than those that lie closer to an edge. 

Since the size of the histogram scales throughout learning, $K(d)$ is also scaled in proportion to the bin size. This is done in each epoch by resetting the width parameter $l = f_{l}\sqrt{(c_x/2)^2+(c_y/2)^2}$ for a fixed fraction $f_{l}$ and with side lengths of a bin $c_x$ and $c_y$.

\section{Scaling Behavior}
\subsection{Scaling of Capacity and Stability}
\begin{figure}
    \centering
    \includegraphics{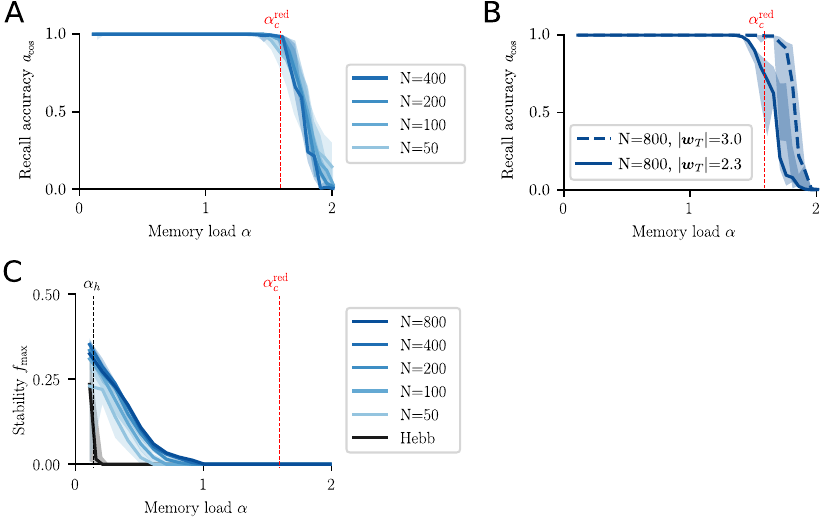}
    \caption{\textbf{Network capacity and stability increases linearly with network size, with minor increases to capacity.} \\
    \textbf{A:} Accuracy of networks trained with the redundancy rule is shown as a function of memory load $\alpha$ for different network sizes $N$. This figure reproduces \autoref{fig:capacity}C with neuron counts $N$ between $50$ and $400$. As the number of neurons increases, the drop in accuracy at $\alpha_c^\mathrm{red}$ becomes sharper. The median over 20 runs is shown.
    \textbf{B:} For $N=800$, scaling the target weights from $2.3$ to $3.0$ leads to retaining high $\alpha_c$. Here, the median over $5$ runs is shown.
    \textbf{C:} The stability $f_\mathrm{max}$ of networks trained on redundancy as a function of memory load $\alpha$. This panel reproduces results for the redundancy goal from \autoref{fig:comparison}C for different network sizes $N$. The shape of the stability curve remains similar as network size is increased.  The median over 20 runs is shown. For $N=400,800$ this was reduced to $5$ runs.
    }
    \label{fig:scaling}
\end{figure}

    One of the fundamental properties of Hopfield networks is that the memory capacity $\alpha_c$ scales linearly with the network size $N$. To support this claim for infomorphic Hopfield networks, we repeated the experiments from \autoref{fig:capacity} and \autoref{fig:comparison} at different network sizes (\autoref{fig:scaling}). Both the memory load $\alpha$ and the stability $f_\mathrm{max}$ are defined as fractions of the network size $N$ and should thus produce the same curve at different $N$. For the memory capacity, we compare at sizes $N \in [50,100,200,400,800]$. For sizes $50-400$, all hyperparameters are unchanged, and the transition around $\alpha_c^\mathrm{red} \approx 1.59$ stays the same. 
    For $N=800$, we repeated the experiments twice with different values for $|\boldsymbol{w}_T| \in \{2.3,3\}$. For $|\boldsymbol{w}_T| = 2.3$, the capacity falls to $\alpha_c \approx 1.41$, while it increases to $\alpha_c \approx 1.66$ for $|\boldsymbol{w}_T| = 3$. These moderate changes indicate that important parameters may have different optimal values at different scales, and that in particular the external target weights have to be adapted to the amount of internal recurrent input a neuron receives.

    We find that stability also scales linearly with network size. There are small increases in $f_\mathrm{max}$ as a function of $\alpha$ across most memory loads. However, only the maximum memory load with $f_\mathrm{max}>0$ keeps increasing with network size. The other increases in $f_\mathrm{max}$ saturate for large networks.
    
\begin{figure}[htbp]
    \centering
    \includegraphics{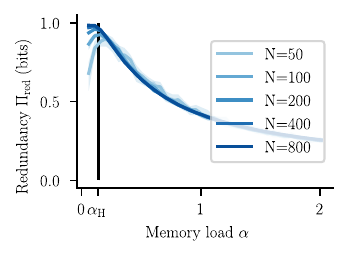}
    \caption{{ \textbf{Redundancy estimates are robust with respect to network size.}
    The redundancy profile from \autoref{fig:hebb}D for Hebbian networks at different network sizes $N$. The curves show the median over $20$ network initializations, with values first averaged across all neurons. For very small memory loads the estimate converges due to finite size effects as explained in Appendix~\autoref{app:finite} (\autoref{fig:finite_size}).}
    }
    \label{fig:hebb_scaling}
\end{figure}

    In addition, the redundancy estimates for Hebbian networks are also independent of network size (see \autoref{fig:hebb_scaling}).

\subsection{Finite Size Effects}
\label{app:finite}
\begin{figure}[t]
    \centering
    \includegraphics{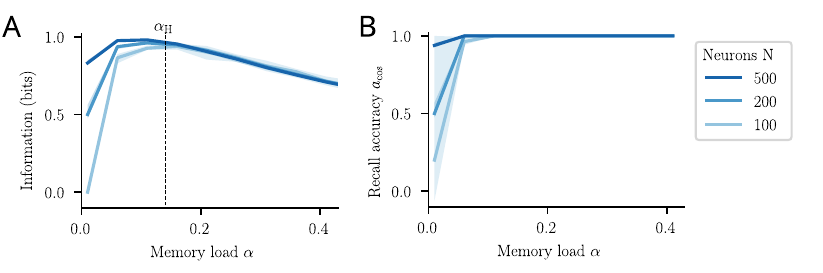}
    \caption{\textbf{Finite size effects for very low memory loads can be counteracted by increasing the network size.} \textbf{A:} The mean redundancy of neurons for different network sizes shows that finite size effects can be reduced for small memory loads $\alpha$ by increasing the number of neurons. For higher memory loads, those effects become smaller.
    \textbf{B:} For networks trained under the redundancy goal, missing redundancy leads to unsuccessful training for small networks and small memory loads $\alpha$.
    In \textbf{A}, values are first averaged across all neurons, and the curve then shows the median of these averages across the initializations. In \textbf{B}, the curve shows the median calculated across $20$ network initializations. For both panels, the shaded area represents the central $90 \%$ of the data.}
    \label{fig:finite_size}
\end{figure}

At small network sizes and very small memory loads, a small portion of neurons will receive constant target inputs with an entropy of $H(T)=0$ and thus $I(Y:T)=0$ and $\Pi_\mathrm{red} = 0$, leading to reduced recall accuracy (\autoref{fig:finite_size}). This effect happens if neurons receive the same target input $T\in\{-w_T,w_T\}$ across all patterns. Writing the absolute number of memory patterns as $m=\alpha N$, and considering uncorrelated patterns, the expected number of neurons that receive only one target label can be written as $\langle N_\mathrm{const}\rangle(\alpha) = \frac{N}{2^{\alpha N-1}}$, which goes to 0 for large $N$.

At finite sizes, however, the number of atoms without any entropy can become significant. 
In particular, for the low memory loads below the Hebbian capacity $\alpha_H$ this leads to an underestimation of the average redundancy of the network. This effect can be reduced by increasing the number of neurons while keeping $\alpha$ fixed (\autoref{fig:finite_size}\textbf{A}). 
This explains why the finite size effects can only be observed for small memory loads and can be countered by increasing the number of neurons as in \autoref{fig:hebb}.
For infomorphic neurons, the finite size effects also prevent learning successfully at small memory loads $\alpha$ in small networks (\autoref{fig:finite_size} \textbf{B}). Because of the higher capacity of infomorphic networks, they are less important at the memory capacity $\alpha_c^\mathrm{red}$, such as in \autoref{fig:capacity}.

\section{Derivation of Redundancy for Associative Memory}
\label{app:HebbianRedundancy}
\subsection{Redundancy as a Necessary Requirement for Associative Memory}
\label{app:HebbianRedundancy_ss1}
    In the following, we provide further intuition for why high redundancy is found in Hebbian Hopfield networks as well as other learning algorithms.

    For successful recall of memories, each neuron's activity needs to be aligned with its respective target value, i.e. $Y=1$ iff $T>0$. Both in the Hebbian and infomorphic Hopfield networks we use a Heaviside activation function in the test phase, implying that $Y=1$ iff $R\geq 0$. The goal of training thus becomes to change the recurrent weights such that the conditional distributions for each target value become disjoint and restricted to one quadrant of the $R$-$T$-plane each, with $p(r\mid T\geq 0)>0$ only if $r>0$, and vice versa $p(r\mid T<0)>0$ only if $r<0$ (see \autoref{fig:visualization}\textbf{A}).
    
    We now show that this separation condition directly implies maximal redundancy. We start from the $I^\mathrm{sx}_\cap$ redundancy measure given in \autoref{eq:Isx}, which can be written as
    \begin{equation}
    \label{eq:isx_repeat}
        \begin{aligned}
        I_\cap^\mathrm{sx}  &= \sum_{r,t,y} p(r,t,y) \log \frac{p(y,r\cup t)}{p(y)p(r\cup t)}\\ 
        &= \sum_{r,t,y} p(r,t,y) \log \frac{p(y,r) + p(y,t) - p(y,r,t)}{p(y)[p(r)+p(t)-p(r,t)]}.
        \end{aligned}
    \end{equation}
    
    We first inspect the second line of \autoref{eq:isx_repeat}. If the two conditional distributions $p(R\mid T=w_T)$ and $p(R\mid T=-w_T)$ are restricted to the first and third quadrant, respectively and the output $Y$ is predicting the target $T$ correctly, then $\forall y,r,t: p(y,r,t) > 0 \implies p(y,r,-t) =  p(y,-r,t) = 0$. As a result, the union probability in the denominator (indicated by a purple cross in \autoref{fig:visualization}) collapses to
    \begin{equation}
        \begin{aligned}
            p(r\cup t) &= p(r) + p(t) - p(r,t)\\
            &= p(r,t) + p(r,-t) + p(t) - p(r,t)\\
            &= p(r\mid -t)p(-t) + p(t)\\
            &= p(t)
        \end{aligned}
    \end{equation}
    for all $(r,t)$ with non-zero probability mass $p(r,t)$. By the same logic, the union probability in the numerator collapses to $p(y, r\cup t) = p(y, t)=p(t)$. Combining these effects leads to redundant information taking on its maximum value given by the entropy of the target signal as
    \begin{align}
        I_\cap^\mathrm{sx}  &= \sum_{r,t,y} p(r,t,y) \log \frac{p(y,r\cup t)}{p(y)p(r\cup t)}\\
        &= \sum_{r,t,y} p(r,t,y) \log \frac{1}{p(y)} = H(Y) = H(T) = 1 \,\mathrm{bit}.
    \end{align}

In the case of imperfect prediction, $p(y\mid t) < 1$ and $I_\cap^\mathrm{sx} = I(Y:T)$. This mutual information is maximized if and only if the neuron always fires when $T>0$ and never fires when $T<0$ or vice versa. Note that information theory is agnostic to signs, such that the case of $p(\sgn R \neq \sgn T) = 1$ also maximizes redundancy. \citet{makkeh2023general} dealt with this case by training a separate decoder, here we preclude it through positive target weights $w_T$.

Next, we show that any deviation from the separated supports necessarily decreases the redundancy. If the target weights $w_T$ are close to zero, as is the case in Hebbian learning, the Heaviside activation function $p(y|r,t)=\delta_{y, \sgn(r+t)}$ reduces to $\delta_{y, \sgn(r)}$ for almost all inputs, which produces 
\begin{equation}
        \begin{aligned}
        I_\cap^\mathrm{sx} 
        &= \sum_{r,t,y} p(r,t,y) \log \frac{p(y,r\cup t)}{p(y)p(r\cup t)}\\
        &= \sum_{r,t,y} \delta_{y,\sgn(r)}p(r,t) \log \frac{p(y,r) + p(y,t) - p(y,r,t)}{p(y)[p(r)+p(t)-p(r,t)]}\\
        &= \sum_{r,t}p(r, t) \log \frac{p(Y = \sgn r,r) + p(Y = \sgn r,t) - p(Y = \sgn r,r,t)}{p(y)[p(r)+p(t)-p(r,t)]}\\
        &= \sum_{r,t}p(r, t) \log \left[\frac{p(r) + \sum_{\sgn r'=\sgn r}p(r',t) - p(r,t)}{p(y)[p(r)+p(t)-p(r,t)]}\right].
        \end{aligned}
    \end{equation}

In general, we have $\sum_{\sgn r'=\sgn r}p(r',t) \leq p(t)$. In particular, this turns into a strict inequality if the conditional probability $p(R\mid T=t)$ has support for both positive and negative values of $r$. In this case, the logarithm will become negative and consequently the redundant information will not be maximal, $I_\cap^\mathrm{sx}<1=H(Y)$. From this, it follows that redundancy can only be maximal when the two conditional distributions $p(R\mid T=t)$ have support on opposite sides of the $R=0$ line, and thus are non-overlapping. Since the disjoint probability mass functions are the only way to achieve perfect recall accuracy, redundancy is therefore necessary for associative memory function.

\subsection{Optimizing Redundancy in Infomorphic Networks}

Having established redundancy as necessary for associative memory function, it remains to be shown that it can also function as a goal to be optimized directly in infomorphic networks.

\begin{figure}[htbp]
    \centering
    \includegraphics[width=\linewidth]{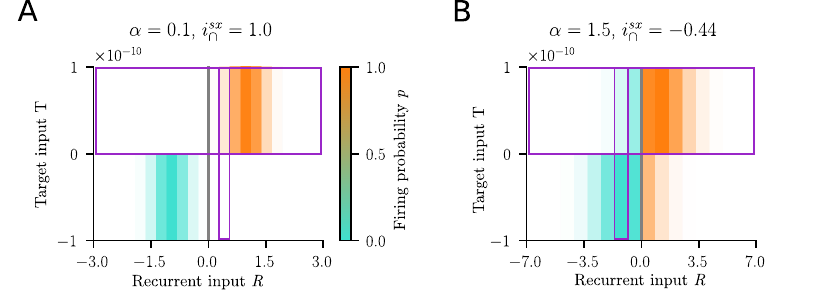}
    \caption{\textbf{Redundancy measures the stability of associative memory solutions.} 
    \textbf{A:} Input distribution for neurons in a Hebbian network trained at a memory load of $\alpha = 0.1$, below the memory capacity $\alpha_H$. The firing probability $p$ is indicated by hue and the frequency of inputs by saturation. The recurrent inputs $R$ are aligned to the sign of $T$ with no misaligned samples despite some variance. 
    The pointwise redundancy considers all bins with \emph{either} the same $R$ \emph{or} the same $T$, represented by the purple cross. Here, redundancy is high across samples, since the distributions neither overlap nor cross the decision boundary indicated in gray.
    \textbf{B:} Input distribution for Hebbian networks trained at a memory load of $\alpha=1.5$, well above the memory capacity. The variance in recurrent inputs $R$ has increased to the point that a relevant fraction of inputs are misaligned w.r.t.\ the target $T$. Redundancy is reduced because for fixed target, the neuron has non-zero probability for both outputs. In addition, for some bins the vertical component of pointwise redundancy also includes samples from the other distribution.
    }
    \label{fig:visualization}
\end{figure}

An important distinction to the Hopfield setup discussed so far is that during training, infomorphic networks use a sigmoid activation function. Thus, the probability of an output $Y$ is dependent on the magnitude of the activation function's inputs $R$ and $T$. For our case of fixed target weights $w_T$, gradient-based learning can only change the recurrent weights and thus move the conditional probability distributions along the r-axis in \autoref{fig:visualization}. Under the sigmoid activation function a global maximum is obtained for the support of $p(R\mid T)$ going to $\pm\infty$. The sigmoid thus provides smooth gradients towards the optimal weight configuration, making gradient-based learning possible.

\begin{figure}[htbp]
   \centering
   \includegraphics[width=0.5\textwidth]{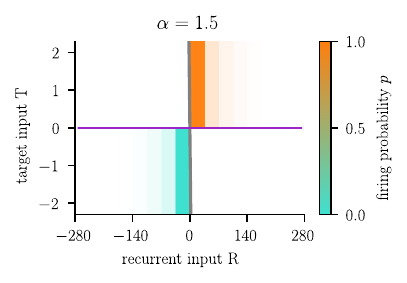}
   \caption{\textbf{Training on redundancy leads to asymmetric recurrent input distributions with disjoint supports.} Joint distribution of inputs $(R,T)$ for a network trained on redundancy at memory load $\alpha = 1.5$. Hue indicates the firing probability $p(Y = +1 \mid R, T)$ and saturation the sample density. The conditional distributions $p(R \mid T = \pm w_T)$ have disjoint supports on opposite sides of $R=0$, in agreement with the geometric condition for maximal redundancy and successful recall (Appendix \ref{app:HebbianRedundancy_ss1}). The shape of the distribution is asymmetric, unlike the symmetric Gaussian distributions found in the Hebbian case.
   }
   \label{fig:gardner}
\end{figure}

In practice, networks trained on redundancy learn to separate the two conditional distributions $p(R\mid T= \pm1)$. Instead of the normal distributions found in Hebbian learning, recurrent inputs $R$ are skewed towards larger absolute values (\autoref{fig:gardner}).

\subsection{Analytical Derivation of Redundancy in Hebbian Hopfield Networks}

Hebbian Hopfield networks in particular allow for an analytical derivation of the redundancy for a given memory load.
 
First, we take the definition of the joint probability distribution $p(y,r,t)=p(t) p(r\mid t) p(y\mid r,t)$, where 
$y\in\{-1,+1\}$ is the neuron output, $r$ the recurrent input and $t\in\{-1\cdot w_T,+1 \cdot w_T\}$ the pattern element. In \autoref{sec:analysis}, the pattern element was defined as $t=w_T\xi_i^{(q)}$. Since $w_T$ only scales the label of the pattern input but does not enter the network dynamics, the joint distribution $p(y,r,t)$ is invariant under the choice of $w_T$. We therefore set $w_T =1$ without loss of generality. Using that $y = \sgn (r)$ is a deterministic function of r, we can rewrite the joint probability distribution as 
\begin{equation}
   p(y,r,t) = \frac{1}{2}\cdot p(r\mid t) \cdot p(y\mid r),
\end{equation}
where $p(y \mid r) = \delta_{y,\sgn (r)}$. So we only have to define $p(r \mid t)$. Following the derivation of \citet{McEliece1987TheCO}, we decompose the local field using the Hebbian weights as defined in \autoref{eq:HebbBinary}. For a network with $N$ neurons storing $m$ patterns and assuming perfect recurrent input $r$ of a pattern $\xi^{(q)}$ ($y_j = \xi^{(q)}_j \ \forall j \neq i$), the local field of a single neuron $i$ becomes
\begin{equation}\label{eq:decomp}
    r_i^{(q)} = \sum_{j\neq i}^N w_{ij}\, \xi_j^{(q)}
    = (N-1)\, \xi^{(q)}_i + \sum_{j\neq i}^N \sum_{{p} \neq q}^m \xi_i^{(p)}\, \xi_j^{(p)}\, \xi_j^{(q)}.
\end{equation}
The assumption holds, if we look at the first iteration of the Hopfield network as done in the main part of the paper.
Each of the $\xi_i^{(p)} \xi_j^{(p)} \xi_j^{(q)}$ in the double sum is a Rademacher random variable (i.e.,\ $\pm1$ with equal probability), and there are $L=(N-1)(m-1)$ such independent terms. If $k$ of the $L$ terms equal $+1$, the sum evaluates to $2k-L$, with $k\sim\mathrm{Binomial}(L,\frac{1}{2})$. Using our definition of the $t$ input ($t = \xi_i^{(q)}$), the conditional distribution can be written as
\begin{equation}
    \label{eq:rcond}
    p\left(r \mid t\right) = \binom{L}{\frac{r-(N-1)\,t+L}{2}}\frac{1}{2^L}.
\end{equation}
It is important to note that $R$ is \emph{not} continuous as it is a sum of $L+1$ terms that are each $\pm1$ with the allowed values being $r\in \{(N-1)\,t-L,(N-1)\,t-L+2,\cdots,(N-1)\,t+L\}$, i.e.\ integers spaced by $2$. With this, we get the discrete joint distribution 
\begin{equation}\label{eq:joint}
    p(y,r,t) = \frac{1}{2} \binom{L}{\frac{r-(N-1) t+L}{2}} \frac{1}{2^L}\delta_{y,\mathrm{sgn}(r)}.
\end{equation}

\begin{figure}[t]
    \centering
    \includegraphics[width=1\linewidth]{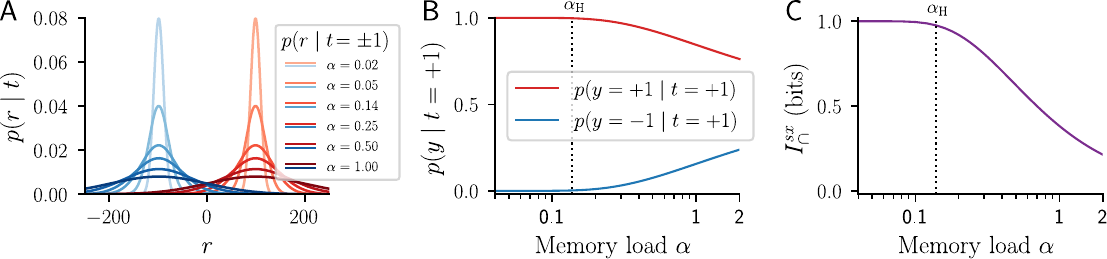}
    \caption{\textbf{Redundant information between target and recurrent input are high at low memory loads and decreases for larger memory loads.} \textbf{A:} Conditional distribution $p(r \mid t \pm 1)$ with $N=100$ neurons at increasing memory loads $\alpha$. The two distributions (red: $t=+1$, blue: $t=-1$) are well separated at low $\alpha$ and progressively overlap near the decision boundary $r=0$ as more patterns are stored. \textbf{B:} Probability of correct ($y=t$ in red) and incorrect ($y\neq t$ in blue) neuron output as a function of $\alpha$. Below the Hopfield capacity ($\alpha_\mathrm{H} \approx 0.138$) the errors are negligible and increase steadily above it. \textbf{C:} Analytically calculated redundancy $I_\cap^\mathrm{sx}(Y:R;T)$. The result agrees with the empirical results from \autoref{sec:analysis}
    }
    \label{fig:AnalyticRedundancy}
\end{figure}

Now, we can use the formulation of $I^\mathrm{sx}_\cap$ derived above given by 
\begin{equation}
    I^\mathrm{sx}_\cap = \sum_{r,t,y} p(r,t,y) \log \frac{p(y\mid t) + p(y,r\mid -t)}{p(y)[1 + p(r\mid -t)]}
\end{equation}
to find an expression for the redundancy. From \autoref{eq:joint} follows directly that $p(y)=\frac{1}{2}$ which simplifies the logarithm to
\begin{equation}
    \log_2 \frac{2(p(r \mid -t)+p(y\mid t))}{p(r\mid -t)+1}
\end{equation}
in all cases of non-zero join probability $p(y,r,t)\neq 0$.

Due to the symmetry of the neurons we can use \\$p(r \mid t=+1)=p(-r\mid t=-1)$ to combine the $t= \pm1$ contributions to derive at
\begin{equation}\label{eq:RedHopf}
    \begin{aligned}
        I_\cap^\mathrm{sx}(Y:R;T) &= \sum_r p(r\mid t = +1) \log_2 \frac{2 \left(p(r \mid t=-1)+p(\sgn (r) \mid t=+1\right)}{p(r\mid t = -1)+1}\\
        &= \sum_r p(r\mid t = +1) \log_2 \frac{2\ p(r \mid t=-1)+2+2p(\sgn (r) \mid t=+1)-2}{p(r\mid t = -1)+1}\\
        &= \sum_r p(r\mid t = +1) \log_2 \left(2 + \frac{2\ p(\sgn (r) \mid t=+1)-2}{p(r\mid t = -1)+1}\right)
    \end{aligned}
\end{equation}
where the sum runs over all allowed values of $r$ as defined above. As the definition of \autoref{eq:RedHopf} depends on the probability $p(r \mid t)$, which is defined based on a discrete grid spacing, the redundancy depends on both $N$ and $m$ separately. 

\autoref{eq:rcond} directly gives us the expression for $p(r\mid t=-1)$ and $p(r\mid t=+1)$ while the probability $p(\sgn(r)\mid t=+1)$ can be written as
\begin{equation}
    \begin{aligned}
        p(\sgn(r)=+1\mid t=+1) &= \sum_{r \geq 0} p(r\mid t=+1) = \sum_{k=k^*}^L \binom{L}{k}\frac{1}{2^L}\\
        p(\sgn(r)=-1\mid t=+1) &= \sum_{r < 0} p(r \mid t=+1) = \sum_{k=0}^{k^*-1} \binom{L}{k}\frac{1}{2^L},
    \end{aligned}
\end{equation}
where $k^*$ is given by $\left\lceil \frac{L-N+1}{2}\right\rceil$ which corresponds to the value of $k$ where $r$ switches sign. The qualitative behavior of the different distributions are shown in \autoref{fig:AnalyticRedundancy}.

\autoref{eq:RedHopf} now analytically shows that redundancy is high if the memory load is low: the distributions $p(r \mid t=+1)$  and $p(r\mid t=-1)$ have essentially no overlap (see \autoref{fig:AnalyticRedundancy}\textbf{A}). So for any $r$ where $p(r\mid t=+1)$ has a significant mass, $p(r\mid t=-1)$  is negligible. At the same time, $p(\sgn (r) \mid t=+1) \approx 1$ since the neuron is almost always correct for low $\alpha$. On the other hand, if the memory load increases, the probability that the neuron fires correctly $p(\sgn(r)\mid t=+1)$,  goes down (see \autoref{fig:AnalyticRedundancy}\textbf{B}) while the overlap between the $p(r \mid t=+1)$  and $p(r\mid t=-1)$ goes up. This results in a decrease of the redundancy for an increase in $\alpha$ (see \autoref{fig:AnalyticRedundancy}\textbf{C}). This result supports the empirical result shown in \autoref{sec:analysis}. 

\subsection{Robustness of PID Estimation}
\label{app:BinningStability}

\begin{figure}[htbp]
    \centering
    \includegraphics{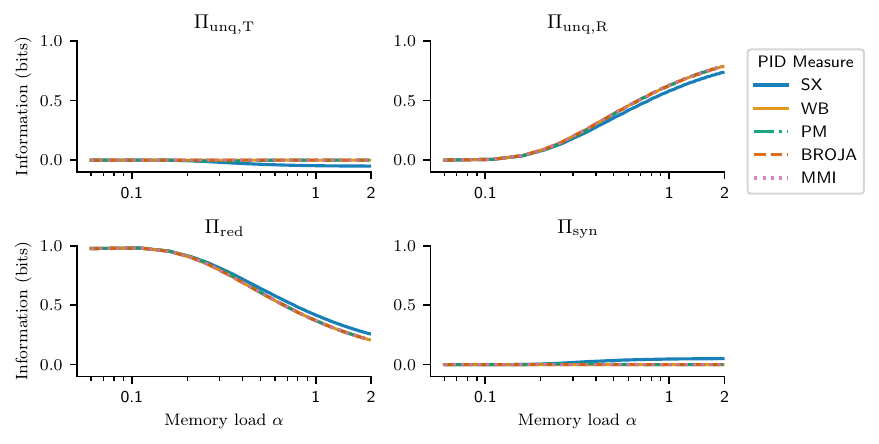}
    \caption{\textbf{The analysis of the PID profile stays consistent between various different measures of PID.} The four panels show the profile of the different information atoms using various PID measures. While there is a numerical difference between the measures, the qualitative result are consistent. The residual entropy is not included as it is no information atom in the classical sense and does not depend on the choice of PID measure. Curves show the median of $20$ network initializations, with values first averaged across all neurons $500$ neurons. Shaded areas indicate the central $90$\% percentile of values.}
    \label{fig:altPIDs}
\end{figure}

\begin{figure}[htbp]
    \centering
    \includegraphics{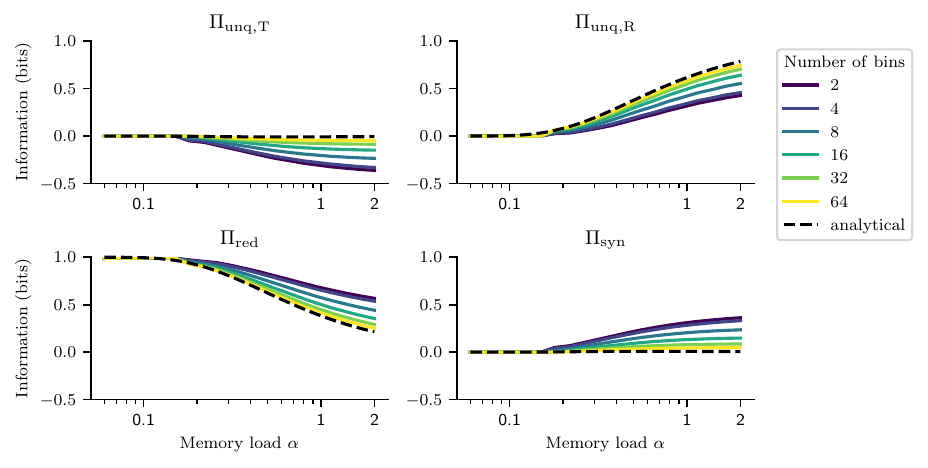}
    \caption{\textbf{The qualitative results of the PID profile of the Hebbian Hopfield network are consistent between different estimations of probability distributions.} The four panels show the information atoms discussed in \autoref{fig:hebb}.\textbf{D}. The different curves show the results for different number of bins used in the dimension of the recurrent input $n_r$ (the standard setting being $n_r = 60$). While the numerical results depend on the number of bins, the qualitative results of an increase of $\Pi_{\unq,R}$ together with a decrease of $\Pi_{\red}$ is consistent among all curves. For larger $n_r$, the results become closer to the analytical results based on the derivation above. Curves show the median of $20$ network initializations, with values first averaged across all $500$ neurons. Shaded areas indicate the central $90$\% percentile of values.}
    \label{fig:n_bins}
\end{figure}
To validate the empirical results of the analysis of Hebbian Hopfield networks discussed in \autoref{sec:analysis} and the analytical results above, which are based on the $I_\cap^\mathrm{sx}$ measure, we performed the same analysis for different PID measures. In particular, we applied $I_\cap^\mathrm{BROJA}$ \cite{bertschinger2014quantifying}, $I_\cap^\mathrm{min}$\cite{williams2010nonnegative}, $I^\mathrm{PM}$\cite{finn2018pointwise} and $I^\mathrm{MMI}$\cite{BertschingerMMI} (see \autoref{fig:altPIDs}). While there are small numerical differences between the results of the different measures, the overall qualitative results agree with the results based on the $I_\cap^\mathrm{sx}$ measure. We used the \textsc{dit} toolbox \cite{dit} to compute all additional measures except $I_\cap^\mathrm{BROJA}$, for which we used the optimized implementation provided by \citet{makkeh2018broja}.

As an additional check of the stability of the estimation in \autoref{sec:analysis}, we also compare the results for different number of bins used in the dimension of the recurrent input $n_r$  (see \autoref{fig:n_bins}). As before, the results are not identical which is due to an estimation bias depending on the number of bins. However, the qualitative results are consistent with the results reported in \autoref{fig:hebb}. With an increase of $n_r$, the empirical results approach the analytical results derived on the calculation above. The choice of $n_r = 60$ for our experiments strikes a balance between computational efficiency and estimation accuracy.

Note that in all of the results, the sample size used to estimate the probability distributions is given by $N \alpha$, which is dependent on the memory load $\alpha$ and the number of neurons $N$. However, the results do not change if we limit the number of samples to be constant for different memory loads.

\section{Correlations}
\label{app:Correlations}

\begin{figure}[hp!]
    \centering
    \includegraphics{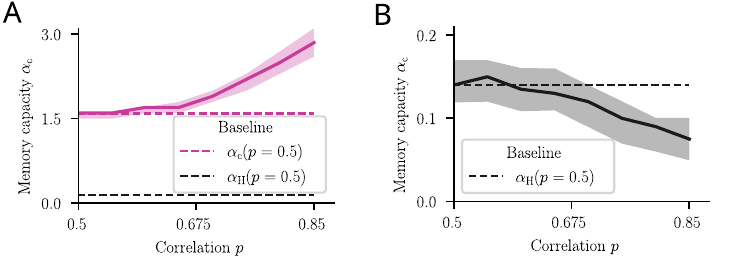}
    \caption{\textbf{Infomorphic Hopfield networks increase their capacity when trained on spatially correlated patterns.}
    \textbf{A:} When spatial correlations $p$ within patterns are introduced, the memory capacity improves for networks trained on the redundancy goal.
    \textbf{B:} Introducing the same correlations as in \textbf{A}, the Hebbian rule is able to encode fewer memories.
    The curves in both panels show the median values across 20 network initializations, and the shaded areas denote the corresponding
    central $90\%$ percentile.}
    \label{fig:correlation}
\end{figure}

The capacity results presented here assume that patterns are created i.i.d. and in particular without spatial correlation. In first experiments, we find that for the redundancy method, more correlation leads to improved capacity (\autoref{fig:correlation}\textbf{A}). This is in contrast to Hebbian learning, where the same correlation decreases capacity (\autoref{fig:correlation}\textbf{B}).

To create spatial correlations, each pattern $\xi^{(p)}$ is procedurally generated in a stepwise manner, where the probability of flipping the next element of a pattern $p(\xi^{(p)}_{i+1}=1 |\xi^{(p)}_{i}=-1 )=p(\xi^{(p)}_{i+1}=-1 |\xi^{(p)}_{i}=1 )p$ is controlled by the parameter $p$. The first element of each pattern is chosen randomly with $p(\xi^{(p)}_1=1)=0.5$. Increasing $p$ increases the average size of blocks within a pattern without changing the balance of $+1$ and $-1$ in patterns. While capacity improves for the redundancy method, further investigation could explore other methods of generating correlated patterns, e.g.\ between patterns, or compare changes in capacity with other learning methods or theoretical bounds.

\section{Ablation Experiments}
\label{App:Ablation}

\begin{figure}
    \centering
    \includegraphics[width=\linewidth]{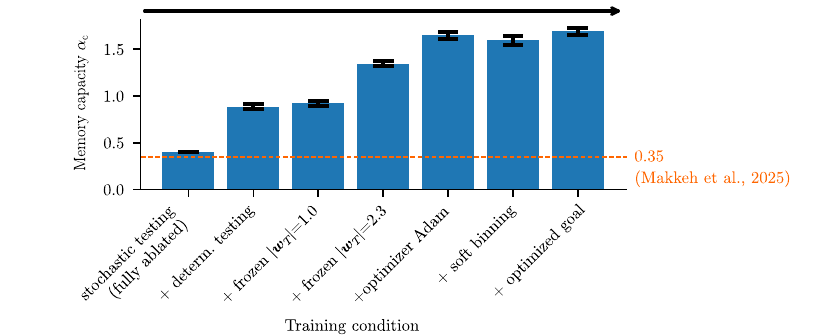}
    \caption{
    \textbf{Ablation experiments point to the most effective features of the training procedure for reaching high memory capacity.} In a sequential approach we removed several features of the training procedure for the left-most condition and added them back in one by one. The conditions are (i) fully ablated, (ii) deterministic testing using a Heaviside activation function (vs. stochastic sigmoidal), (iii) frozen target weights $\boldsymbol{w}_T$ (vs. learnable ones), (iv) higher initial scale of target weights (2.3 vs 1.0), (v) improved optimizer for training (Adam vs. SGD), (vi) binning method (differentiable soft vs. hard) and (vii) optimized goal (vs. pure redundancy maximization). The fully ablated condition reaches a capacity comparable to the one reported in \cite{makkeh2023general}. The error bars correspond to the standard deviation over 20 runs. 
    }
    \label{fig:ablation}
\end{figure}

Our infomorphic Hopfield networks build on those introduced in \cite{makkeh2023general}, but employ multiple changes to the network and training procedure. Through these changes, the resulting memory capacity increased more than fourfold, from 0.35 to above 1.59. To quantify the effect of the most relevant changes, we conducted a series of ablation experiments. Starting from a `fully ablated' condition that most closely mirrors the setup of \cite{makkeh2023general}, features of the training procedure are successively added in while measuring their effect on memory capacity (\autoref{fig:ablation}).

The five features varied in the ablation are:
\begin{itemize}
    \item \textbf{Testing procedure:} stochastic neuron outputs over 20 recall time steps (as in \citet{makkeh2023general}) → deterministic outputs with up to 100 recall time steps.
    \item \textbf{Target weights $|\boldsymbol{w}_T|$:} learnable, initialized at scale $1.0$  → frozen at scale $1.0$ → frozen at scale $2.3$.
    \item \textbf{Optimizer:} SGD → Adam.
    \item \textbf{PID estimation:} hard (non-differentiable) binning → soft binning (see \autoref{app:Binning}).
    \item \textbf{Goal function:} $G = \Pi_\mathrm{red}$ → optimized goal $G^{(i)}$ from \autoref{sec:optRedRule}.
\end{itemize}
Out of these, the most effective changes are the deterministic testing procedure, the fixed scale of target weights of $2.3$ and the switch to the more sophisticated Adam optimizer.

Further differences to the original were omitted from the ablation study because they were not needed to match the 0.35 capacity reported in \cite{makkeh2023general}. First, the goal function was simplified from $G = \Pi_\mathrm{red} + 0.1\, I(Y; R, T)$ in \cite{makkeh2023general} to pure redundancy, $G = \Pi_\mathrm{red}$, while retaining high capacity and stable learning. Second, the activation function was simplified from
$A(r, t) = \frac{r^8 \, \mathrm{sign}(r) + t^8 \, \mathrm{sign}(t)}{r^8 + t^8} \cdot \left(r^8 + t^8\right)^{1/8}$
to the linear form $A(r, t) = r + t$. Additionally, we note that the target weights in \cite{makkeh2023general} differ from our setup in ways beyond initialization scale: there, they are drawn from a normal distribution (with both positive and negative values). The Hopfield layer was also followed by a simple decoder layer to compensate for inverse neuron representations. These structural differences are outside the scope of the ablation.

\end{document}